\documentclass[12pt,a4paper]{article}
\pdfoutput=1
\usepackage{epsfig}
\usepackage{mathrsfs}
\usepackage{graphics}
\usepackage{amssymb}

\usepackage{tikz}
\usepackage{comment}
\usetikzlibrary{decorations.pathmorphing}

\usepackage{amsmath}
\usepackage{amsthm}
\usepackage{amsfonts}

\def\hybrid{\topmargin -20pt    \oddsidemargin 0pt
        \headheight 0pt \headsep 0pt
        \textwidth 6.25in       % A4 paper
        \textheight 9.5in       % A4 paper
        \marginparwidth .875in
        \parskip 5pt plus 1pt   \jot = 1.5ex}

\hybrid

\def\baselinestretch{1.2}

\catcode`\@=11

\def\marginnote#1{}
\newcount\hour
\newcount\minute
\newtoks\amorpm
\hour=\time\divide\hour by60
\minute=\time{\multiply\hour by60 \global\advance\minute by-\hour}
\edef\standardtime{{\ifnum\hour<12 \global\amorpm={am}%
        \else\global\amorpm={pm}\advance\hour by-12 \fi
        \ifnum\hour=0 \hour=12 \fi
        \number\hour:\ifnum\minute<10 0\fi\number\minute\the\amorpm}}
\edef\militarytime{\number\hour:\ifnum\minute<10 0\fi\number\minute}
\def\draftlabel#1{{\@bsphack\if@filesw {\let\thepage\relax
   \xdef\@gtempa{\write\@auxout{\string
      \newlabel{#1}{{\@currentlabel}{\thepage}}}}}\@gtempa
   \if@nobreak \ifvmode\nobreak\fi\fi\fi\@esphack}
        \gdef\@eqnlabel{#1}}
\def\@eqnlabel{}
\def\@vacuum{}
\def\draftmarginnote#1{\marginpar{\raggedright\scriptsize\tt#1}}

\def\draft{\oddsidemargin -.5truein
        \def\@oddfoot{\sl preliminary draft \hfil
        \rm\thepage\hfil\sl\today\quad\militarytime}
        \let\@evenfoot\@oddfoot \overfullrule 3pt
        \let\label=\draftlabel
        \let\marginnote=\draftmarginnote
   \def\@eqnnum{(\theequation)\rlap{\kern\marginparsep\tt\@eqnlabel}%
\global\let\@eqnlabel\@vacuum}  }

\def\preprint{\twocolumn\sloppy\flushbottom\parindent 2em
        \leftmargini 2em\leftmarginv .5em\leftmarginvi .5em
        \oddsidemargin -.5in    \evensidemargin -.5in
        \columnsep .4in \footheight 0pt
        \textwidth 10.in        \topmargin  -.4in
        \headheight 12pt \topskip .4in
        \textheight 6.9in \footskip 0pt
        \def\@oddhead{\thepage\hfil\addtocounter{page}{1}\thepage}
        \let\@evenhead\@oddhead \def\@oddfoot{} \def\@evenfoot{} }

\def\numberbysection{\@addtoreset{equation}{section}
        \def\theequation{\thesection.\arabic{equation}}}

\def\underline#1{\relax\ifmmode\@@underline#1\else
        $\@@underline{\hbox{#1}}$\relax\fi}

\def\titlepage{\@restonecolfalse\if@twocolumn\@restonecoltrue\onecolumn
     \else \newpage \fi \thispagestyle{empty}\c@page\z@
        \def\thefootnote{\fnsymbol{footnote}} }

\def\endtitlepage{\if@restonecol\twocolumn \else \newpage \fi
        \def\thefootnote{\arabic{footnote}}
        \setcounter{footnote}{0}}  %\c@footnote\z@ }

\catcode`@=12
\relax

\def\figcap{\section*{Figure Captions\markboth
        {FIGURECAPTIONS}{FIGURECAPTIONS}}\list
        {Figure \arabic{enumi}:\hfill}{\settowidth\labelwidth{Figure
999:}
        \leftmargin\labelwidth
        \advance\leftmargin\labelsep\usecounter{enumi}}}
 \relax
\def\tablecap{\section*{Table Captions\markboth
        {TABLECAPTIONS}{TABLECAPTIONS}}\list
        {Table \arabic{enumi}:\hfill}{\settowidth\labelwidth{Table
999:}
        \leftmargin\labelwidth
        \advance\leftmargin\labelsep\usecounter{enumi}}}
 \relax
\def\reflist{\section*{References\markboth
        {REFLIST}{REFLIST}}\list
        {[\arabic{enumi}]\hfill}{\settowidth\labelwidth{[999]}
        \leftmargin\labelwidth
        \advance\leftmargin\labelsep\usecounter{enumi}}}
 \relax
\makeatletter
\newcounter{pubctr}
\def\publist{\@ifnextchar[{\@publist}{\@@publist}}
\def\@publist[#1]{\list
        {[\arabic{pubctr}]\hfill}{\settowidth\labelwidth{[999]}
        \leftmargin\labelwidth
        \advance\leftmargin\labelsep
        \@nmbrlisttrue\def\@listctr{pubctr}
        \setcounter{pubctr}{#1}\addtocounter{pubctr}{-1}}}
\def\@@publist{\list
        {[\arabic{pubctr}]\hfill}{\settowidth\labelwidth{[999]}
        \leftmargin\labelwidth
        \advance\leftmargin\labelsep
        \@nmbrlisttrue\def\@listctr{pubctr}}}
 \relax
\makeatother
\newskip\humongous \humongous=0pt plus 1000pt minus 1000pt

\newif\ifdtup

\relax

\def\be{\begin{equation}}
\def\ee{\end{equation}}
\def\ba{\begin{eqnarray}}
\def\ea{\end{eqnarray}}

\def\no{\noindent}

\def\bl{\bigl}
\def\br{\bigr}

\def\IR{\relax{\rm I\kern-.18em R}}

\def\co{{\cal O}}

\def\0{{\sst{(0)}}}
\def\1{{\sst{(1)}}}
\def\2{{\sst{(2)}}}
\def\3{{\sst{(3)}}}
\def\4{{\sst{(4)}}}
\def\5{{\sst{(5)}}}
\def\6{{\sst{(6)}}}
\def\7{{\sst{(7)}}}
\def\8{{\sst{(8)}}}
\def\n{{\sst{(n)}}}

 \let\m=\mu \let\n=\nu  \let\r=\rho
 \let\t=\tau    
  \let\D=\Delta  
   \let\F=\Phi

 \def\bd{\begin{document}} \def\ed{\end{document}}
\def\ds{\documentstyle} \let\fr=\frac \let\bl=\bigl \let\br=\bigr
\let\Br=\Bigr \let\Bl=\Bigl
\let\bm=\bibitem
\let\na=\nabla
\let\pa=\partial \let\ov=\overline
\def\ft#1#2{{\textstyle{{\scriptstyle #1}\over {\scriptstyle #2}}}}
\def\fft#1#2{{#1 \over #2}}
\def\del{\partial}
\def\sst#1{{\scriptscriptstyle #1}}
 \def\oneone{\rlap 1\mkern4mu{\rm l}}
\def\ie{{\it i.e.\ }}
\def\via{{\it via}}
\def\semi{{\ltimes}}
\def\str{{\rm str}}
\def\Dm{{{D_{\sst{max}}}}}
\def\vac{ \left | 0 \right \rangle }
\def\kvac{ \left | k \right \rangle }

\def\sp{\; \; \;}

\def\bol{ \left | B (p^+) \right \rangle}
\def\bo1{ \left | B^0 (p^+) \right \rangle}

\def\bolt{ \left | B (p^+) \right \rangle_{\t}}

\def\boxl{ \left | B (x^-) \right \rangle}

\def\<{ \langle }
\def\>{ \rangle }

\def\vf{\varphi}

\def\ls{{(l,0)}}
\def\lv{{(l,\pm1)}}
\def\lt{{(l,\pm2)}}

\def\lse#1{{(l_{#1},0)}}
\def\lve#1{{(l_{#1},\pm1)}}
\def\lte#1{{(l_{#1},\pm2)}}

\def\lsg#1{{5(l_{#1},0)}}
\def\lvg#1{{5(l_{#1},\pm1)}}
\def\ltg#1{{5(l_{#1},\pm2)}}

\def\lsi#1{{5{(#1,0)}}}
\def\lvi#1{{5{(#1,\pm1)}}}
\def\lti#1{{5{(#1,\pm2)}}}

\def\lsr#1{{1{(#1,0)}}}
\def\lvr#1{{1{(#1,\pm1)}}}
\def\ltr#1{{1{(#1,\pm2)}}}

\def\cn{{\cal N}}
\def\cao{{\cal O}}
\def\cD{{\cal D}}
\def\cE{{\cal E}}
\def\cF{{\cal F}}
\def\cG{{\cal G}}
\def\cH{{\cal H}}
\def\cK{{\cal K}}
\def\cO{{\cal O}}
\def\cP{{\cal P}}
\def\cQ{{\cal Q}}
\def\cR{{\cal R}}
\def\cS{{\cal S}}
\def\cT{{\cal T}}
\def\cU{{\cal U}}
\def\cV{{\cal V}}
\def\cW{{\cal W}}
\newcommand{\nono}{\nonumber}
\newcommand{\dtilde}[1]{\tilde{\tilde{#1}}}
\newcommand{\hatb}[1]{\hat{\ov{#1}}}
\newcommand{\hatt}[1]{\hat{\tilde{#1}}}
\newcommand{\emnr}{{e_\m}^{\n\r}}

\begin{document}
%\draft
%$\mathscr{abcdefghijklmnopqrstuvwxyz}$
%\renewcommand{\theequation}{\arabic{equation}}
\renewcommand{\theequation}{\thesection.\arabic{equation}}

\newcommand{\beq}{\begin{equation}}
\newcommand{\eeq}[1]{\label{#1}\end{equation}}
\newcommand{\ber}{\begin{eqnarray}}
\newcommand{\eer}[1]{\label{#1}\end{eqnarray}}
\newcommand{\eqn}[1]{(\ref{#1})}
\begin{titlepage}
\begin{center}

\hfill April 2014\\

\vskip .4in

{\large \bf Non-equilibrium dynamics and $AdS_4$ Robinson-Trautman}

\vskip 0.6in

{\bf Ioannis Bakas$^{a)}$ and Kostas Skenderis$^{b)}$}
\vskip 0.2in
{\em a) Department of Physics, School of Applied Mathematics and Physical Sciences \\
National Technical University, 15780 Athens, Greece\\
b) Mathematical Sciences and STAG Research Center, University of Southampton \\
Highfield, Southampton SO17 1BJ, United Kingdom \\
\vskip 0.2in
\footnotesize{\tt bakas@mail.ntua.gr, k.skenderis@soton.ac.uk}}\\

\end{center}

\vskip .8in

\centerline{\bf Abstract}

\no
The Robinson-Trautman space-times provide solutions of Einstein's equations with negative cosmological
constant, which settle to $AdS_4$ Schwarzschild black hole at late times. Via gauge/gravity duality they should
describe a system out of equilibrium that evolves towards thermalization. We show that the area of the past apparent horizon
of these space-times satisfies a generalized Penrose inequality and we formulate as well as provide evidence for a
suitable generalization of Thorne's hoop conjecture. We also compute the holographic energy-momentum tensor and
deduce its late time behavior. It turns out that the complete non-equilibrium process on the boundary is governed
by Calabi's flow on $S^2$.  Upon linearization, only special modes that arise as supersymmetric zero energy states
of an associated supersymmetric quantum mechanics problem contribute to the solution. We find that
each pole of radiation has an effective viscosity given by the eigenvalues of the Laplace operator on $S^2$ and there
is an apparent violation of the KSS bound on $\eta / s$ for the low lying harmonics of large $AdS_4$
black holes. These modes, however, do not satisfy Dirichlet boundary conditions, they are out-going
and they do not appear to have a Kruskal
extension across the future horizon ${\cal H}^+$.

\vfill
\end{titlepage}
\eject

\def\baselinestretch{1.2}
\baselineskip 16 pt
\noindent

\tableofcontents

%\newpage

\section{Introduction}
\setcounter{equation}{0}

An important question in physics is to understand the evolution of out-equilibrium quantum systems.
Gauge/gravity duality provides a new tool to study non-equilibrium dynamics, mapping
this problem to classical gravity in AdS space. More precisely, thermal states of the dual theory are mapped
to AdS black holes and the process of thermalization is mapped to black hole formation. Close to thermal equilibrium,
the long wave-length, late time behavior of QFT is described by hydrodynamics and, furthermore, this regime has
a corresponding bulk description in terms of solutions constructed in a gradient expansion.
The complete description of non-equilibrium phenomena require global time-dependent solutions approaching the
(appropriate) AdS black hole at late times. With this in mind, much effort has been devoted in recent years in obtaining
numerical solutions that address the problem in general terms (see, for instance, \cite{Chesler:2013lia} and
references therein). Here, we aim to discuss certain aspects of non-equilibrium dynamics using analytic solutions.
In particular, we focus on a class of gravitational solutions provided by the Robinson-Trautman metrics, which may be
regarded as non-linear version of a specific type of perturbations of $AdS_4$ Schwarzschild black holes.

Linear perturbations of the Schwarzschild black hole (with or without cosmological constant) have been a topic of
immense study over the years, reducing their description to an effective Schr\"odinger problem. What is perhaps
less known is the remarkable fact that the effective quantum mechanics of all four-dimensional black holes is
supersymmetric in that the Hamiltonian can be written as the square of a supercharge. However,
the boundary conditions imposed by the black hole make the Hamiltonian non-Hermitian and its spectrum is not even real.
The allowed energy levels provide the quasi-normal modes of the equilibrium state, which turns out to be stable under all kind of
perturbations. Under appropriate boundary conditions, the effective supersymmetry pairs the non-zero modes: parity
even (polar) perturbations are paired with parity odd (axial) perturbation and, as usual with supersymmetry, the
zero modes are special satisfying a first order equation. These supersymmetric zero energy modes, which are called
algebraically special modes and they correspond to purely dissipative perturbations of black holes, are the focus of our
work. They can be lifted to exact solutions of the full non-linear Einstein equations, giving rise to the class of
Robinson-Trautman metrics. In that respect, our discussion extends known results to space-times with non-zero cosmological
constant.

When the cosmological constant is zero, the solutions we are considering represent isolated gravitationally radiating
systems. The system relaxes to equilibrium by radiating the excess energy, which escapes to null infinity, and settles
to a spherically symmetric configuration provided by the Schwarzschild solution.
One can quantify the energy that is radiated away by computing the corresponding Bondi mass, \cite{BondiMass}:
as energy escapes to infinity, the Bondi mass decreases monotonically (in retarded time) and becomes equal to the
Schwarzschild mass after infinitely long time. The space-time has a naked singularity in the past, reflecting the fact
that the solution does not include the sources generating the excitation. There is, however, a past apparent horizon
which may be regarded as a surface that surrounds the isolated system. When the cosmological constant is negative,
there are new issues because the asymptotic structure of space-time is different. Since asymptotically AdS space-times
do not have null infinity, the radiation cannot escape. Instead, it is absorbed at the boundary modifying the boundary
metric.

In AdS/CFT correspondence, the boundary conditions represent sources for dual operators and
the boundary metric is the source for the energy-momentum tensor of the dual theory. The energy-momentum tensor is
one of main observables that one wishes to extract and use it to study the approach to thermal equilibrium.
The boundary geometry of $AdS_4$ Robinson-Trautman space-times is topologically $\mathbb{R} \times S^2$,
with $\mathbb{R}$ representing time, and the complete non-equilibrium process is governed by the Calabi flow on $S^2$.
This flow is a well-known geometric evolution equation in mathematics describing volume preserving deformations of
K\"ahler manifolds ($S^2$ in our case) given by a fourth order non-linear diffusion process. Starting from a general
initial metric on $S^2$, the flow continuously deforms the metric transforming it to the constant curvature metric after
infinitely long time. The Robinson-Trautman solution provides a space-time interpretation of the two-dimensional
Calabi flow with retarded time playing the role of the deformation parameter. While other geometric evolution equations,
such as the normalized Ricci-flow are also related to linear perturbations of large $AdS_4$ black holes, as will be seen later,
only the Calabi flow is known to be associated with exact non-linear solutions of Einstein equations for all values of the
cosmological constant.

In the context of gauge/gravity duality, one expects to find that sufficiently close to equilibrium the energy-momentum
tensor takes the form of dissipative hydrodynamics. We will indeed find that this is the case. One should note, however,
that the underlying physics is very different than in other discussions that appeared in the literature so far.
In most works, the following problem is possed: one creates a non-equilibrium state by coupling the dual QFT to an
external source for a finite time interval. After switching off the interaction the system loses energy to the black
hole and equilibrates. The physical mechanism for dissipation is black hole absorption. In our case, however, there is
only out-going radiation and nothing is absorbed by the black hole; instead, the dissipation is due to the external couplings
of the system. The fact that the radiation is out-going means that the perturbation ought to vanish at the horizon, as it does.
Expressing the out-going modes in terms of an in-coming basis, one finds that these modes are not smooth at the horizon,
in general, and for sufficiently low multi-poles of radiation there is no Kruskal extension at all. While this is
in principle a major concern, one may view the solutions as representing the gravitational field outside an isolated compact
object losing its asymmetry via gravitational radiation until it relaxes to a spherical symmetric state.
In that picture, the complete solution should be described by a matter-filled interior solution joined to the
Robinson-Trautman solution, which is now covering only the exterior region, but it remains an open problem how
precisely to implement it.

In standard treatments of holography, the linearized perturbations around a given solution satisfy Dirichlet boundary
conditions and in-coming boundary conditions at the horizon yield two-point functions in the dual QFT. From the low energy
limit of those correlation functions one subsequently extracts the transport coefficients via the Kubo formula.
In our case, the algebraically special modes satisfy mixed boundary conditions, due to the first order equation they obey,
which in turn imply unusual boundary conditions for the metric and, moreover, these modes are purely out-going.
Thus, a clean holographic interpretation of these modes is somewhat challenging. We find that the effective viscosity depends
on the eigenvalues of the Laplacian on $S^2$ (related to the pole of radiation we are considering) and for
sufficiently low harmonics of large $AdS_4$ black holes there is a violation of the KSS bound on $\eta/s$.
It should be said right away that
non-universality and violation of the KSS bound have also been observed in higher derivative theories of gravity
\cite{higherDer} as well as in two-derivative gravity coupled to matter in the presence of anisotropy \cite{Aniso}
(see, for instance, \cite{Cremonini:2011iq} for a topical review and further references), but the underlying physics is
different in those studies compared to ours.

The Bondi mass, which is naturally defined at null infinity of asymptotically flat space-times, describes the total
mass of the system at some given instant of (retarded) time. For $AdS_4$ Robinson-Trautman there is no null infinity,
but one can still define by analogy a ``Bondi mass" that decreases monotonically under the evolution and eventually
becomes equal to the Schwarzschild mass, as in the asymptotically flat case. Moreover, we will show that it
satisfies a Penrose inequality, which is a central condition for the formation of horizons in general relativity.
It turns out that the Bondi mass is greater than a specific function of the area of the past
apparent horizon, hereby extending known results for the Robinson-Trautman metrics with $\Lambda = 0$.
A closely related condition for the formation of horizons is Thorne's hoop conjecture stating that the mass $M$ gets
compacted into a region whose circumference is bounded in all directions. We will also provide evidence for a suitable
version of this conjecture that holds for Robinson-Trautman space-times with negative cosmological constant, using the Bondi
mass and some appropriately chosen circumference to surround its past apparent horizon. The two inequalities are
independent from each other, in general, though one follows from the other under some special conditions that are also
discussed in the text.

This paper is organized as follows. In section 2, we introduce the Robinson-Trautman solutions, as spherical gravitational
waves, and show that they are non-linear versions of special ``supersymmetric" perturbations of $AdS_4$ Schwarzschild metric.
We also present their relation to Calabi flow and extract their late time behavior. In section 3, we discuss a number of
global issues, focusing on the (lack of) Kruskal extension and the existence, uniqueness and explicit construction of the
past apparent horizon. We prove a Penrose inequality for the Bondi mass and formulate as well as provide evidence for
Thorne's hoop conjecture in the presence of negative cosmological constant. In section 4, we consider holographic aspects
of the solutions by computing the corresponding energy-momentum tensor. We determine its late time behavior and compute
the effective viscosity of the system. Here, we also define an entropy current of the boundary fluid and show by 
explicit computation in late time expansion that there is entropy production upon thermalization. 
In section 5, we present our conclusions and outline some open problems.
We also collect a number of useful results in three appendices. In Appendix A, we review the derivation of the
Robinson-Trautman solution for the benefit of the readers. In Appendix B, we review the relevant aspects of large and
small $AdS_4$ black holes and collect some useful formulae that are implicitly used in the text. Finally, in Appendix C,
we review the theory of polar perturbations of spherical black holes.

In all expressions appearing in the text, Newton's constant $G$ is normalized to 1, whereas the gravitational coupling constant
$8 \pi G$ is denoted by $\kappa^2$, as usual.

\newpage

\section{Spherical gravitational waves}
\setcounter{equation}{0}

The Robinson-Trautman metrics provide an exact class of radiative solutions of Einstein equations,
which are available for all values of the cosmological constant $\Lambda$, including, in particular,
$\Lambda < 0$. This class exists in four space-time dimensions
and has been thought to describe the effect of spherical gravitational waves
emitted by bounded sources. Although such metrics do not capture the general features of gravitational
radiation, they are often interesting to consider in detail at linear and non-linear level.
Higher dimensional generalizations of Robinson-Trautman metrics have also been considered in the literature,
but they do not seem to support non-trivial radiative solutions. Here, we summarize some basic facts about them
that will be used later.

\subsection{Robinson-Trautman metrics}

The Robinson-Trautman metrics are singled out in four space-time dimensions by their unique geometric
property to admit a geodesic null congruence with zero shear and twist and non-vanishing divergence. Thus,
the metric has the following general form, \cite{robi} (but see also the textbooks
\cite{kramer, carmeli, griffiths} and some early references on the subject \cite{ted, lukacs, schmidt}),
\be
ds^2 = 2r^2 e^{\Phi (z, \bar{z}; u)} dz d\bar{z} - 2dudr - F(r, u, z, \bar{z})
du^2 ~.
\ee
The variable $r$ is a radial coordinate in space
and $u$ is a {\em retarded time} coordinate. Closed
surfaces with constant $r$ and $u$ represent distorted two-dimensional
spheres in a system of conformally flat (K\"ahler) coordinates
$(z, \bar{z})$.

Using Einstein equations, in the presence of a cosmological constant $\Lambda$, the
form of the front factor $F$ is uniquely determined in terms of $\Phi$ as
\be
F = r\partial_u \Phi
- \Delta \Phi - {2m \over r} - {\Lambda \over 3}r^2 ~,
\ee
where
\be
\Delta = e^{-\Phi} \partial_z \partial_{\bar{z}} = {1 \over 2} \nabla^2
\ee
denotes the Laplacian on the distorted 2-spheres. The cosmological constant term in $F$
accounts for the appropriate asymptotic behavior of the metric as $r \rightarrow \infty$.
The parameter $m$ is an integration constant, which is taken to be positive and it is related to
the physical mass of the configuration. The remaining Einstein equations are completely solved,
provided that $\Phi (z, \bar{z}; u)$ satisfies the following
parabolic fourth order non-linear differential equation on $S^2$,
\be
3m \partial_u \Phi + \Delta \Delta \Phi = 0 ~,
\label{basiceq}
\ee
which is called Robinson-Trautman equation. Note that this equation is insensitive to the presence of $\Lambda$.
The essential steps of the derivation are summarized for completeness in Appendix A.

General $u$-dependent solutions are thought to describe spherical gravitational waves, in a sense that will
be made more precise later,
in the full non-linear regime of Einstein equations. Although equation \eqn{basiceq} has been studied
extensively in physics as well as in mathematics (for it arises in a seemingly unrelated problem in geometric
analysis, namely the Calabi flow on $S^2$, as well be seen later in more detail),
hardly any non-trivial explicit solutions exist to this day. However, some strong qualitative results are
available in the literature regarding the long time existence of the general solution. For appropriate (smooth)
initial data, the $u$-dependent solutions converge towards a fixed point,
as $u \rightarrow \infty$, \cite{schmidt, rendall, piotr1}, which exhibits stability against small perturbations.

The fixed point is the $u$-independent solution of the Robinson-Trautman equation
associated to the round metric of radius $1$ on $S^2$ with conformal factor
\be
e^{\Phi_0} = {1 \over \left(1 + z \bar{z}/2  \right)^2} ~.
\ee
The corresponding four-dimensional space-time metric reads
\be
ds^2 = {2r^2 \over \left(1 + z \bar{z}/2 \right)^2}
dz d\bar{z} - 2dudr - \left(1-{2m \over r} - {\Lambda \over 3}
r^2 \right) du^2
\ee
and it coincides with the Schwarzschild solution in the Eddington-Filkenstein
frame in the presence of cosmological constant $\Lambda$. Indeed, by introducing the change of coordinates
\be
u=t-r_{\star} ~, ~~~~~ z = \sqrt{2} ~ {\rm cot} {\theta \over 2} ~
e^{i\phi} ~,
\ee
where $r_{\star}$ is the so called tortoise coordinate of the corresponding Schwarzschild
metric, which is detailed in Appendix B together with other useful definitions, we arrive at the standard
form of the solution
\be
ds^2 = -f(r)dt^2 + {dr^2 \over f(r)} + r^2 \left(d\theta^2 +
{\rm sin}^2 \theta d\phi^2 \right)
\ee
with the required profile function
\be
f(r) = 1 - {2m \over r} - {\Lambda \over 3} r^2 ~.
\ee

\subsection{Linearization of the solutions}

The Robinson-Trautman equation can be regarded as non-linear diffusion
process (of fourth order)
on $S^2$, which has the topology of the black hole horizon. Since small
curvature perturbations of the round sphere correspond to linear perturbations
of the Schwarzschild metric as the configuration tends towards
equilibrium as $u \rightarrow \infty$, it is appropriate to
parametrize the fluctuating two-dimensional metric as
\be
ds_2^2 = [1 + \epsilon_l (u) P_l ({\rm cos} \theta)]
\left(d\theta^2 + {\rm sin}^2 \theta d\phi^2 \right)
\label{defo}
\ee
with small parameters $\epsilon_l (u)$. Here, we only consider axially symmetric deformations,
without loss of generality, using Legendre polynomials $P_l (\xi)$ with $l \geq 2$.
The cases $l=0$ and $l=1$ are omitted, since we are paying attention
to quadrupole radiation terms or higher on physical grounds. This parametrization
preserves the area $A$ of $S^2$ to linear order in $\epsilon_l$, since for all $l \geq 1$ we have
\be
\delta_{\epsilon} A = {(-1)^l \over 2} V_0 ~ \epsilon_l (u)
\int_{-1}^{+1} d\xi ~ P_l (\xi) = 0 ~.
\ee

To work out the form of the Robinson-Trautman equation to linear order, we set for convenience
\be
K(\theta, u) = \epsilon_l(u) P_l ({\rm cos} \theta) ~.
\ee
A simple calculation shows that
\be
3m \partial_u K = - \Delta_0 [(\Delta_0 +1) K] ~,
\ee
where $\Delta_0$ denotes the Laplacian of the round $S^2$ whose eigenfunctions
$P_l ({\rm cos} \theta)$ have eigenvalues $-l(l+1)/2$; likewise, the corresponding eigenvalues of the
operator $\Delta_0 + 1$ are $-(l-1)(l+2)/2$.
This immediately yields the following evolution for
$\epsilon_l (u)$, \cite{ted},
\be
\epsilon_l (u) = \epsilon_l (0) {\rm exp} \left(-{u \over 12m}
(l-1)l(l+1)(l+2) \right) \equiv \epsilon_l (0) e^{-i \omega_{\rm s} u} ~,
\ee
where $\omega_{\rm s}$ denote collectively the set of purely imaginary frequencies (damping modes)
\be
\omega_{\rm s} = -i {(l-1)l(l+1)(l+2) \over 12m}
\ee
with $l \geq 2$. The spectrum of $\omega_{\rm s}$ does not depend on the
cosmological constant $\Lambda$, as for the Robinson-Trautman equation.

The result, which approximates well the asymptotic behavior of the
full non-linear evolution as $u \rightarrow \infty$, shows that
all curvature perturbations are damped exponentially fast until the
configuration finally takes its canonical round metric form. All such
linear solutions represent different modes of gravitational radiation
in multi-pole expansion, when elevated to the four-dimensional
space-time metric. Higher order poles of radiation are damped faster
than the lower ones, as expected on physical grounds.
In more detail, using small $u$-dependent perturbations of the form
\eqn{defo}, the Robinson-Trautman line element takes the following form
in the vicinity of the Schwarzschild solution,
\ba
ds^2 & = & - \left(1 - {2m \over r} - {\Lambda \over 3} r^2
+ r \partial_u K - (\Delta_0 + 1)K \right) du^2
-2 dudr  \nonumber \\
& & + r^2 (1 + K) \left(d\theta^2 + {\rm sin}^2 \theta d\phi^2
\right) ~,
\ea
choosing to work with spherical coordinates $(\theta, \phi)$ and
using $K(\theta, u)$ as defined above.
Then, taking into account the dependence of $\epsilon_l$ upon $u$ and
the action of $\Delta_0$ on $K$,
it turns out that
\be
r \partial_u K - (\Delta_0 + 1) K = {1 \over 2}(l-1)(l+2)
\left(1-{r \over 6m}l(l+1)\right) K ~.
\ee

Changing variables $(u, r)$ to $(t, r)$ by $u = t-r_{\star}$,
so that direct comparison can be subsequently made with the canonical form of
gravitational perturbations of black holes, it follows that the
underlying metric takes the form
\be
g_{\mu \nu} = g_{\mu \nu}^{(0)} + \delta g_{\mu \nu}
\ee
around the static Schwarzschild metric, which is assigned the superscript $(0)$.
The perturbations are tabulated in terms of the
$(t, r, \theta, \phi)$ components, as
\be
\delta g_{\mu \nu} = \left(\begin{array}{cccc}
f(r)H(r) & -H(r) & 0 & 0 \\
  &   &   &   \\
-H(r) & H(r)/f(r) & 0 & 0 \\
  &   &   &   \\
0 & 0 & r^2 & 0 \\
  &   &   &   \\
0 & 0 & 0 & r^2 {\rm sin}^2 \theta
\end{array} \right)
e^{-i\omega_{\rm s} (t-r_{\star})}
P_l ({\rm cos} \theta) ~,
\label{ident1}
\ee
where
\be
H(r) = {i\omega_{\rm s} \over f(r)}
\left(r - {6m \over l(l+1)} \right) .
\label{ident2}
\ee
The integration constant $\epsilon_l (0)$ represents the freedom to scale
the form of the linear perturbations $\delta g_{\mu \nu}$ by an overall factor.

These perturbations are the so called algebraically special modes of the black hole,
studied by Chandrasekhar \cite{chandra3} and connected
to Robinson-Trautman metrics, upon linearization, for $\Lambda = 0$ \cite{schutz} (but see also \cite{couch} 
for earlier work on the subject).
Here, we are simply extending all that to $\Lambda \neq 0$, as in \cite{reall1}.
These modes are special in many respects as we will discuss shortly and
the preceding analysis shows that the class of Robinson-Trautman metrics
are formed by a consistent non-linear superposition of them.
Note that it does not  seem possible to truncate the Robinson-Trautman equation to solutions that encompass
a finite number of algebraically special modes, for, otherwise, we would have simple models
of gravitational radiation in our disposal (though at late times this looks plausible).
More general radiative metrics are formed by non-linear superposition of all quasi-normal modes,
not just the algebraically special ones, and hence they are notoriously more difficult to study,
though physically more interesting.

The algebraically special modes are special case of the polar perturbations of
spherical black holes. The theory of perturbations of spherical black holes is a well studied subject
\cite{wheeler, zerilli, chandra2, kokko, lemos, moss, Bakas:2008gz}. We now briefly summarize some of
the relevant aspects of this theory; more details can be found in Appendix C.
The perturbations can be split into parity even (polar) and parity odd (axial)
perturbations. It turns out that the study of these perturbations can be reduced to the effective Schr\"{o}dinger problems
\be
H_{\pm} \Psi_{\pm} (r_{\star}) = E \Psi_{\pm} (r_{\star}) ~,
\ee
where
\be
H_{\pm}= -{d^2 \over dr_{\star}^2} + W^2 \pm {dW \over dr_{\star}} ~.
\ee
The plus sign corresponds to polar perturbations and the minus to axial perturbations. The function $W$ is given by
\be
W(r) = {6m f(r) \over r[(l-1)(l+2)r + 6m]} + i \omega_{\rm s}
\ee
and
\be
E = \omega^2 - \omega_{\rm s}^2 ~.
\ee
Remarkably, despite the fact that the space-time itself has no supersymmetry and there are not even fermions,
the effective Schr\"odinger problems can be combined as partners into supersymmetric quantum mechanics
with $W$ playing the role of the superpotential. Indeed, as discussed in Appendix C, defining
${\cal H}={\rm diag}(H_+, H_-)$ and $\Psi=(\Psi_+, \Psi_-)^T$, one finds that the two Schr\"{o}dinger problems combine into one
\be
{\cal H} \Psi = E \Psi
\ee
and the Hamiltonian is a square of a supercharge, i.e., ${\cal H} = \{{\cal Q} , ~ {\cal Q}^{\dagger} \}$,
where ${\cal Q}$ is defined as in equation (\ref{Qdef}).
Due to boundary conditions, the Hamiltonian is only formally Hermitian and the spectrum is not even real.

The algebraically special modes are the zero energy states of this problem and they are annihilated by the supercharge
${\cal Q}$. Using equations (\ref{Qdef}) and (\ref{qdef}), we find that these perturbations satisfy the first order equation,
\be \label{susy}
Q \Psi_{+}^{(0)} (r_{\star}) = \left(- {d \over dr_{\star}} + W (r_{\star}) \right)
\Psi_{+}^{(0)} (r_{\star})= 0 ~.
\ee
This equation can be integrated to yield the corresponding wave-function
\be
\Psi_{+}^{(0)} (r_{\star}) = C ~ {r \over (l-1)(l+2) r + 6m} {\rm exp}
\left(i \omega_{\rm s} r_{\star} \right) \, ,
\ee
where $C$ is an integration constant, which, for later convenience, is chosen to be
\be
C= {6m \over l(l+1)}.
\ee
Then, in the notation of Appendix C, we find that the system of equations for
the metric coefficients of polar perturbations admit the following simple solution at
$\omega = \omega_{\rm s}$,
\be
H_0(r) = H_2(r) = -H_1(r) = i {\omega_{\rm s} \over f(r)}
\left(r - {6m \over l(l+1)} \right) K(r)
\ee
and
\be
K(r) =
{\rm exp}\left(i \omega_{\rm s} r_{\star} \right)
= {\rm exp} \left({r_{\star} \over 12m} (l-1)l(l+1)(l+2)
\right)
\ee
that yield precisely $\delta g_{\mu \nu}$ for the linearization problem of the Robinson-Trautman metrics
around the Schwarzschild solution, as given above, for all values of the cosmological constant.

Note that $\Psi_+^{(0)} (r_{\star})$ vanishes on the black hole horizon, irrespective of $\Lambda$, complying
with the physical boundary condition that nothing can come out directly of a
black hole. As $r \rightarrow \infty$, $\Psi_+^{(0)}(r_{\star})$ blows up exponentially when $\Lambda = 0$,
but when $\Lambda < 0$, which is the case of primary interest here, it reaches a finite value.
Thus, for $AdS_4$ black holes, all algebraically special modes have normalizable
wave functions, i.e.,
\be
\int_{-\infty}^{0} dr_{\star} \mid \Psi_{+}^{(0)} (r_{\star}) \mid^2
< \infty ~.
\ee
Furthermore, the algebraically special modes satisfy {\em mixed boundary conditions}
at $r = \infty$.  Indeed, using (\ref{susy}) and evaluating $W(r_{\star})$ at $r = \infty$ (equivalently at $r_{\star} = 0$) we find
\be
{d \over dr_{\star}} \Psi_{+}^{(0)} (r_{\star}) \mid_{r_{\star} = 0} ~ =
\left(i \omega_{\rm s} - {2m \Lambda \over (l-1)(l+2)} \right)
\Psi_{+}^{(0)} (r_{\star} = 0) ~.
\ee
These non-standard boundary conditions are responsible for some peculiar properties of the solution that will be derived later.

\subsection{Calabi flow on $S^2$}

The Robinson-Trautman equation also arises in mathematics and coincides with
the so called Calabi flow on $S^2$, as was first pointed out by Tod \cite{Tod}, which
describes certain type of geometric deformations of the metric
\be
ds_2^2 = 2 e^{\Phi (z, \bar{z}; u)} dz d\bar{z} ~.
\label{coclasa}
\ee
Recall that the Calabi flow is generally defined for metrics $g_{a \bar{b}}$
on K\"ahler manifolds $M$ of
arbitrary dimension with local coordinates $(z^a, \bar{z}^a)$ and assumes
the form \cite{calabi} (but see also \cite{futaki, tian})
\be
\partial_u g_{a \bar{b}}
= {\partial^2 R \over \partial z^a \partial \bar{z}^b}
\ee
in terms of the Ricci curvature scalar $R$ of the metric.
As such, it provides volume preserving deformations within a given K\"ahler
class of the metric. Specializing to $S^2$, we have $R = -2 \Delta \Phi$ and
the identification with the Robinson-Trautman equation follows
immediately, setting $3m = 2$ in appropriate units.

The Calabi flow is a parabolic equation of fourth order, and, as such, it is difficult to
study by standard techniques that are mainly applicable to second order equations. By considering
appropriate (smooth) initial data $g_{a \bar{b}} (z, \bar{z}; 0)$, the flow deforms the metric towards
its canonical form, which is typically represented by a constant curvature metric (provided that the
K\"ahler manifold admits one). The quadratic curvature functional on $M$, also known
as Calabi functional,
\be
{\cal C} = \int_M R^2 \, ,
\ee
where $R$ the curvature scalar, decreases monotonically along that flow, reaching its extremum
at the fixed point (also called {\em extremal metric} on $M$). The fixed point is the familiar constant
curvature metric when $M \simeq S^2$. Many aspects of the
Robinson-Trautman metrics are derived from the mathematical
theory of Calabi flows. The results are qualitatively similar to the
heat equation on $S^2$, thinking of the Ricci scalar curvature $R$ as
temperature; in that case, the average temperature (viz. the
Euler number) remains constant throughout the evolution, whereas the
average temperature square (viz. the quadratic curvature functional)
decreases monotonically until thermal equilibrium is reached at the
fixed point, where the temperature is constant everywhere. This framework has also been
used more generally, as tool, to explore the geometrization of K\"ahler manifolds.

The Calabi flow is one of many geometric evolution equations studied in
mathematics. For later use, we introduce here another geometric evolution equation, known as Ricci flow,
which is of second order and deforms the metric on a Riemannian manifold by the Ricci curvature tensor.
In that case, the volume of space is not preserved under the Ricci flow, but it is always possible to define
a variant of it, known as normalized Ricci flow, which is volume preserving and will be used for comparison
with Calabi flow. The Ricci flow for the
class of conformally flat metrics on $S^2$, \eqn{coclasa}, takes the
following form
\be
\partial_u \Phi = \Delta \Phi ~,
\ee
whereas the corresponding normalized Ricci flow on $S^2$ with fixed area
$4 \pi$ is given by
\be
\partial_u \Phi = \Delta \Phi + 1  ~.
\ee
The constant curvature metric provides the fixed point for both Calabi and
normalized Ricci flow equations on $S^2$. In either case, the canonical
metric is reached from a given initial data after sufficiently long time.

It is instructive to examine the spectrum of linear perturbations around
the equilibrium state, using axially symmetric deformations of the
round sphere parametrized by $\epsilon_l (u) P_l ({\rm cos} \theta)$,
as in the linear approximation \eqn{defo} of the Robinson-Trautman equation.
It can be easily verified that as $u \rightarrow \infty$, the metric perturbations decay as follows,
under the normalized Ricci flow,
\be
\epsilon_l (u) = \epsilon_l (0) {\rm exp} \left(-{u \over 2}  (l-1)(l+2) \right)
\equiv \epsilon_l (0) e^{-i \Omega_{\rm s}} ~,
\ee
with spectrum of purely imaginary frequencies (damping modes) for all $l \geq 2$
\be
\Omega_{\rm s} \sim - i {(l-1)(l+2) \over 2} ~.
\ee
$\Omega_{\rm s}$ is unique up to a universal factor that depends on the
physical scale of the problem.
Later, in section 4, the scale of the Ricci flow will be identified with
$3r_{\rm h}/2$ to match the values of some purely dissipative modes in the
hydrodynamic description of very large $AdS_4$ black holes, whereas the scale of
$u$ of the Calabi flow is provided by $3m$.
The dependence of $\Omega_{\rm s}$ upon $l$, being quadratic versus the quartic dependence of
$\omega_{\rm s}$ for the Calabi flow, reflects the order of the corresponding geometric evolution
equations.

A schematic representation of curvature perturbations on $S^2$ is depicted in Fig.1.
All modes of perturbation dissipate faster under Calabi flow when the radius of the sphere is
sufficiently small. For spheres of large radius, however, the situation is partially reversed, as perturbations
with sufficiently small $l$ dissipate faster under the normalized Ricci flow, whereas modes with large enough
$l$ are still dissipating faster under Calabi flow.

\begin{figure}[h]
\vspace{-9.5cm}\hspace{4cm}\includegraphics[scale=0.5]{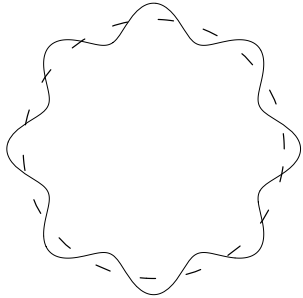}
%\centering \epsfxsize=6cm\epsfbox{F1.eps}
%\vspace{1.cm}
\caption{Curvature perturbations of the round sphere.}
\end{figure}

\subsection{Late-time behavior of solutions}

Next, we describe the late time behavior of solutions to the Robinson-Trautman (Calabi flow) equation
including non-linear effects.
Let us parametrize the conformal factor of the $S^2$ line element as
\be
e^{\Phi (z, \bar{z}; u)} = {1 \over \sigma^2 (z, \bar{z}; u)
\left(1 +
z \bar{z}/2 \right)^2} ~,
\ee
so that $\sigma (z, \bar{z}; u)$ describes the deviations away from
the static black hole solution. First, it was established that
for any sufficiently smooth initial data $\sigma (z, \bar{z}; u_0)$
solutions of the Robinson-Trautman equation exist for all
$u \geq u_0$, \cite{rendall}; the short time existence of solutions was established
earlier using the parabolic nature of the equation in $u$, \cite{schmidt}.
Second, it was found that there is an asymptotic expansion of
$\sigma (z, \bar{z}; u)$ for large $u$ of the following form, \cite{piotr2},
\be
\sigma (z, \bar{z}; u) = 1 + \sum_{p \geq 1, ~ q \geq 0}
\sigma_{p, q} (z, \bar{z})
u^q e^{-2pu/m} ~,
\ee
where $\sigma_{p, q} (z, \bar{z})$ are appropriately chosen
smooth functions on $S^2$ with $\sigma_{0, 0}$ normalized to 1
without loss of generality.
Thus, by exponential damping, the Robinson-Trautman metrics
approach asymptotically fast the Schwarzschild solution,
$\sigma = 1$, as $u \rightarrow \infty$. These results are
quite general and they are valid irrespective of $\Lambda$ for the
Robinson-Trautman equation is insensitive to the presence of cosmological
constant.

The above series expansion of $\sigma (z, \bar{z}; u)$
captures the linear as well as the non-linear effects of gravitational
radiation at late retarded times. Linear effects are solely described
by the system of algebraically special modes, which take the values
\be
i \omega_{\rm s} = {2 \over m} ~, ~~ {10 \over m} ~, ~~
{30 \over m} ~, ~ \cdots
\ee
for $l=2, ~ 3, ~ 4,  \cdots$, respectively. They only contribute to the
terms $\sigma_{1, 0}, ~ \sigma_{5, 0}, ~ \sigma_{15, 0},  \cdots$,
which in this case are described by the corresponding spherical
harmonics on $S^2$, depending on $l$.
Other coefficients in the series are naturally related to non-linear
gravitational effects; they include contributions to all terms
with $p > 1$ as well as to terms with $q \neq 0$.

Although some coefficients $\sigma_{p, q} (z, \bar{z})$ may vanish,
it has already been established that there exist generic solutions with
non-vanishing coefficient $\sigma_{15, 1} (z, \bar{z})$ - the first
possible such term with $q \neq 0$ - so that
the asymptotic expansion assumes the following form,
\ba
\sigma (z, \bar{z}; u) & = &
1 + \sigma_{1, 0}(z, \bar{z}) e^{-2u/m} +
\sigma_{2, 0} (z, \bar{z}) e^{-4u/m}
\cdots + \sigma_{14, 0}(z, \bar{z}) e^{-28u/m} \nonumber\\
& & + [ \sigma_{15, 0}(z, \bar{z}) + \sigma_{15, 1}
(z, \bar{z}) u ]
e^{-30u/m} + {\cal O}\left(e^{-32u/m} \right) .
\ea
Of course, there might be higher order terms
$p \geq 16$ with $q \neq 0$, but their effect is subdominant.

Following Chru\'sciel and collaborators, \cite{piotr2},
we consider axially symmetric deformations,
so that all coefficients depend on the angle $\theta$ and not $\phi$, and
further impose {\em antipodal symmetry} for simplicity. It implies that only even values of
$l$ contribute to the general solution.
Then, in that case, the first few coefficients in the series
turn out to be
\ba
\sigma_{1, 0} (x) & = & a \left(x^2 - {1 \over 3} \right) , \\
\sigma_{2, 0} (x) & = & -a^2 \left({23 \over 78} x^4 - {47 \over 39} x^2
+ {49 \over 234} \right) , \\
\sigma_{3, 0} (x) & = & a^3 \left({997 \over 5226} x^6 -{36697 \over 47034} x^4
+{25309 \over 15678} x^2 - {8899 \over 47034} \right)
\ea
and so on, where $x = {\rm cos} \theta$. The functions above are uniquely determined and they
can be rewritten in terms of the even Legendre polynomials
$P_2 (x) =  (3x^2 - 1)/2$, $P_4 (x)  =  (35x^4 - 30 x^2 + 3)/8$
and so on.

The first term $\sigma_{1, 0}(x)$ represents the quadrupole gravitational
radiation, as in the linear approximation, and is set equal to
$(2a/3) P_2(x)$ using an arbitrary constant $a$. All other terms up to
$\sigma_{5, 0}(x)$ represent non-linear effects of
gravitational radiation. One could add to $\sigma_{5, 0}(x)$ a
term proportional to $P_3 (x)$, with a new coefficient independent of $a$,
which represents the effect of eighth-pole radiation in the linear
approximation and shows up at order ${\exp}(-10u/m)$. However, such
term will be excluded, if antipodal symmetry is imposed.
Higher order terms up to $\sigma_{15, 0}(x)$ are also uniquely
determined in terms of $a$ and they also represent non-linear
effects of quadrupole radiation. One can also add to $\sigma_{15,0}(x)$
a term proportional to $P_4 (x)$ that represents the effect of
sixteenth-pole radiation in the linear approximation and shows up
at order ${\exp}(-30u/m)$. Such a term is compatible with the antipodal
symmetry. There is an additional non-linear effect that appears to this order
attributed to the presence of $\sigma_{15,1}(x)$, which is relatively very small but definitely
not zero; it was determined numerically in terms of $a$ to be
$\sigma_{15, 1} (x) \simeq (0.2155750672866a)^{15} P_4 (x)$,
\cite{piotr2}.
%\be
%\sigma_{15, 1} (x) \simeq (0.2155750672866a)^{15} P_4 (x) ~.
%\ee
Compared to the typical coefficients of $\sigma_{p, 0}(x)$ with
$1 \leq p \leq 15$, $\sigma_{15, 1} (x)$ is smaller by nine orders of magnitude.
The pattern repeats itself for higher values of $p$ although it has not been
studied exhaustively by numerical methods to the best of our knowledge.

The late time expansion of solutions to the Robinson-Trautman equation has implications
for the global structure of the corresponding space-times when considering the Kruskal
extension across the horizon ${\cal H}^+$. It will also be used to provide the late time expansion
of the associated holographic energy-momentum tensor that is constructed later in section 4.

\section{Global aspects}
\setcounter{equation}{0}

The global structure of Robinson-Trautman space-time is sensitive to the
cosmological constant $\Lambda$ (see \cite{piotr2} for $\Lambda = 0$, \cite{dolsky} for 
$\Lambda > 0$, \cite{bicak} for $\Lambda < 0$). For $\Lambda < 0$, it also turns out to depend on the
relative size of $AdS_4$ black holes. This might look surprising at first sight for the
the Robinson-Trautman equation depends only on $m$ and not on $\Lambda$.

\subsection{Kruskal extension}

Let us introduce Kruskal-type coordinates in the bulk space-time,
which are generally defined as
\be
\tilde{u} = -e^{-u/2 \delta_{\rm h}} ~, ~~~~~
\tilde{v} = e^{v/2 \delta_{\rm h}} ~,
\ee
using the retarded and advanced time coordinates, $u = t-r_{\star}$
and $v = t+r_{\star}$, respectively. The parameter $\delta_{\rm h}$
is defined for all values of $\Lambda$, which is taken here to be non-positive, as
\be
\delta_{\rm h} = {r_{\rm h}^2 \over 2(3m - r_{\rm h})}
~,
\ee
using the radius $r_{\rm h}$ of the horizon of the static solution,
and it reduces to the familiar choice $\delta_{\rm h} = 2m$
when $\Lambda = 0$; otherwise, it varies in
the interval $2m > \delta_{\rm h} > 0$ as $\Lambda$
changes from 0 to $-\infty$, keeping $m$ fixed. The surface gravity $\kappa_{\rm gr}$ and the temperature
$T$ of the equilibrium black hole state are related to $\delta_{\rm h}$ as follows,
\be
\kappa_{\rm gr} = {f^{\prime} (r_{\rm h}) \over 2} = 2 \pi T = {1 \over 2 \delta_{\rm h}} ~.
\ee

In these coordinates, the line element of the Robinson-Trautman metric becomes
\be
ds^2 = 2r^2 e^{\Phi} dz d \bar{z} - 4 \delta_{\rm h}^2 f(r)
e^{-r_{\star} / \delta_{\rm h}} d\tilde{u} d\tilde{v} -
4 \delta_{\rm h}^2 (F-f(r)) {d\tilde{u}^2 \over \tilde{u}^2} ~,
\label{crus}
\ee
where
\be
F-f(r) = r \partial_u \Phi - \Delta \Phi -1 ~.
\ee
The static $AdS_4$ black hole has $F=f(r)$ and always admits smooth extension
across the event horizon. Time-dependent solutions, however, have $F \neq f(r)$
and the regularity of the Kruskal extension is questionable.
Are the extensions through the null hypersurface $\mathscr{H}^{+}$
given by $u = \infty$ (equivalently $\tilde{u} = 0$) smooth
or not? If not, it will imply, in particular, that an observer can determine
by local measurements whether he/she has crossed the event horizon.

It turns out that the asymptotic series
expansion of the function $\sigma (z, \bar{z}; u)$, which describes the deviation
of the metric from the canonical form, as $\Phi = \Phi_0 - 2 {\rm log} \sigma$, takes
the following form in terms of the Kruskal coordinate $\tilde{u}$, \cite{piotr2},
\ba
\sigma(z, \bar{z}; \tilde{u}) & = &
1 + \sigma_{1, 0}(z, \bar{z}) (-\tilde{u})^{4 \delta_{\rm h}/m}
+ \sigma_{2, 0}(z, \bar{z}) (- \tilde{u})^{8 \delta_{\rm h} / m} + \cdots
+ \sigma_{14, 0}(z, \bar{z})
(- \tilde{u})^{56 \delta_{\rm h} / m} \nonumber\\
& & + [\sigma_{15, 0}(z, \bar{z}) - 2 \delta_{\rm h}
\sigma_{15, 1} (z, \bar{z})
{\rm log} \mid \tilde{u} \mid ]
(- \tilde{u})^{60 \delta_{\rm h} / m} +
{\cal O} \left((-\tilde{u})^{64 \delta_{\rm h} / m}\right) .
\ea
Due to the presence of the ${\rm log} | \tilde{u} |$
term, the function $\sigma$ is not smooth at $\tilde{u} = 0$
but it is at most $C^{[a]}$-differentiable with $[a]$ being the integer
part of
\be
a = {60 \delta_{\rm h} \over m} - 1 ~.
\ee
In turn, it implies that the Robinson-Trautman metric, which is now written
in the form \eqn{crus}, is at most $C^{[a]-2}$-differentiable
at $\tilde{u} = 0$, because the coefficient in front of the
$d \tilde{u}^2$ term contains an additional factor $1/\tilde{u}^2$.
As will be seen shortly, the regularity of the metric may
be even lower, depending on the size of $\delta_{\rm h}$.
As a result, the extension of the Robinson-Trautman
metric across the horizon $\mathscr{H}^{+}$ can not be
smooth in general.
The degree of differentiability depends crucially on $\Lambda$.
When $\Lambda =0$, $\delta_{\rm h} = 2m$ and $a = 119$,
i.e., the full four-dimensional metric is only
$C^{117}$-differentiable and not $C^{\infty}$, \cite{piotr2}. When
$\Lambda < 0$, which is the case of primary interest here, the degree of
differentiability decreases simply because $\delta_{\rm h} < 2m$.
In fact, the parameters $m$ and $\Lambda$ can be easily arranged so that
the metric is not even $C^1$; it may also very well be that
$\partial \sigma / \partial \tilde{u}$ diverges at $\mathscr{H}^{+}$.

Actually, there is a striking difference between $\Lambda = 0$
and $\Lambda < 0$ black holes, which is already seen at the linear level.
Retaining only the contribution of the algebraically special modes
in the asymptotic expansion of $\sigma (z, \bar{z}; \tilde{u})$, one
observes that $F-f(r)$ is of order
${\cal O} (\tilde{u}^{4 \delta_{\rm h} / m})$ as $\tilde{u}$ approaches
0, receiving the most dominant contribution from the quadrupole radiation term.
As a result, the overall coefficient of the $d \tilde{u}^2$ term of the
metric is of order ${\cal O}(\tilde{u}^{(4 \delta_{\rm h}/ m) - 2})$.
Thus, for $\Lambda = 0$ black holes, this coefficient is of order
${\cal O}(\tilde{u}^6)$ and the extension through the horizon appears
to be smooth at the linear level, as noticed in
reference \cite{Tod}; of course, we know better now that the smoothness
breaks down due to non-linear effects attributed to the presence of the
non-vanishing term $\sigma_{15, 1} (z, \bar{z})$, \cite{piotr2}. On the other hand, for
$\Lambda < 0$ black holes, it may happen that $(4 \delta_{\rm h}/m) - 2$
is negative, in which case there are divergences at $\mathscr{H}^{+}$
already appearing at the linear level, \cite{bicak}. For example, one may
consider large $AdS_4$ black holes with $m > r_{\rm h}$, \cite{don}, which actually have
$4 \delta_{\rm h} / m < 1$, as can be easily verified. By the same token,
small AdS black holes have $4 \delta_{\rm h} / m > 1$ and in fact the
lower bound exceeds the value 2 for sufficiently small (i.e., not of
intermediate size) black holes. For the relevant definitions of large versus small black holes
we refer the reader to Appendix B.

Thus, the Kruskal extension of the Robinson-Trautman metric breaks down
completely for large $AdS_4$ black holes because the coefficient of
$d \tilde{u}^2$ becomes singular at $\tilde{u} = 0$. This is induced
perturbatively by the first algebraically special mode (associated to
quadrupole radiation) and signals an instability towards the formation of a null singularity
at $\mathscr{H}^{+}$. It is intriguing that this perturbative instability appears only for large black holes.
Note that in the original work \cite{bicak} the authors did not describe the result in thermodynamic terms,
distinguishing large from small AdS black holes, \cite{don}.

More generally, within the linear approximation, each algebraically
special mode gives a contribution to the $d \tilde{u}^2$
term of the Kruskal extension of the metric of order
${\cal O}(\tilde{u}^{2i \omega_{\rm s} \delta_{\rm h} - 2})$, since
\be
F-f(r) = {(l-1)(l+2) \over 2} \left(1 - {l(l+1) \over 6m} r \right)
P_l({\rm cos} \theta) \left(- \tilde{u} \right)^{2i \omega_{\rm s}
\delta_{\rm h}} \, .
\ee
Thus, for large black holes and sufficiently low $l$ satisfying
the bound\footnote{Note that the $(tt)$ and $(rr)$ components of the algebraically special
perturbations $\delta g_{\mu \nu}$ of large $AdS_4$ black holes become negative close to
$r_{\rm h}$ for precisely those values of $l$ (see equations \eqn{ident1} and \eqn{ident2}).}
\be
{m \over r_{\rm h}} > {l(l+1) \over 6} ~,
\ee
we have the relation
\be
2i \omega_{\rm s} \delta_{\rm h} = {(l-1)l(l+1)(l+2) r_{\rm h}^2 \over
12m (3m - r_{\rm h})} < 1
\ee
generalizing the effect of the quadrupole term to all other potentially
dangerous terms arising in the multi-pole expansion. For higher values of $l$
we have $2i \omega_{\rm s} \delta_{\rm h} > 1$, but the lower bound may still
not exceed the value 2. It can also be verified that large black holes
with sufficiently large values of $l$,
so that
\be
\sqrt{2} ~ {m \over r_{\rm h}} < {l(l+1) \over 6} ~ ,
\ee
always satisfy the higher bound $2i \omega_{\rm s} \delta_{\rm h} > 2$. It is rather
odd behavior of large $AdS_4$ black holes, which are favored thermodynamically, \cite{don}.
Sufficiently small black holes, on the other hand, satisfy the bound
$2i \omega_{\rm s} \delta_{\rm h} > 2$ for all values of $l$.

One should note, however, that the curvature
invariants are smooth at ${\cal H}^+$. In fact, all invariants formed by powers of the  Riemann tensor (without taking
covariant derivatives) have the same value as the invariants evaluated on the Schwarzschild solution; this was shown
in reference \cite{rendall} when $\Lambda = 0$ and we checked it for the Kretschmann invariant when $\Lambda <0$.
On the other hand, the authors of reference \cite{piotr2} argued that at least one component of the Riemann tensor
(or any power of covariant derivatives acting on it)  will blow up at ${\cal H}^+$ in any coordinate system.
This indicates that there is a null singularity at the future horizon.

Summarizing, the Kruskal extension of the Robinson-Trautman metric breaks
down completely for $AdS_4$ black holes when the coefficient of
the $d \tilde{u}^2$ term becomes singular at $\tilde{u} = 0$.
The radiative solutions settle down to $AdS_4$ Schwarzschild black hole
at late retarded times, and, sometimes, depending on $l$,
they can be extended across the horizon to include the black hole
interiors. In those cases, the interior of a static black hole
can be jointed to an external Robinson-Trautman space-time across
$\mathscr{H}^{+}$ located at $\tilde{u} = 0$, but the extension is not smooth.
The Penrose diagram in Fig.2 does not include the extension across the horizon.
The situation resembles the behavior of out-going modes of a scalar field in $AdS_4$ black hole
background, which are not smooth near the future horizon, as pointed out in reference \cite{horohube}.
Fortunately, the smoothness of $\mathscr{I}$ at $r = \infty$ is not affected at all by possible
discontinuities across the horizon.

\begin{figure}[h]
\hspace{3.5cm}\includegraphics[scale=1.5]{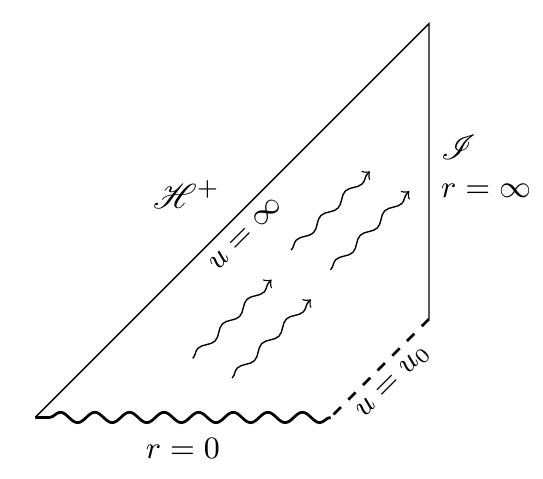}
\caption{Penrose diagram of Robinson-Trautman
space-times with $\Lambda < 0$.}
\end{figure}

\subsection{Past apparent horizon}

The Robinson-Trautman space-time exhibits a past apparent horizon. The
construction we present here generalizes previously known results for
$\Lambda = 0$, \cite{Tod}, to the case $\Lambda < 0$.

Let us define a 2-surface $\Sigma$ in space-time by the following
embedding relations,
\be
u = {\rm const.} ~, ~~~~~ r = U (z, \bar{z}) ~.
\ee
The unit tangent vectors to $\Sigma$ with $(u, ~ r, ~ z, ~ \bar{z})$
components are
\be
m^{\mu} = {1 \over re^{\Phi / 2}} \left(0, ~ \bar{\partial} U , ~
0, ~ 1 \right) ~, ~~~~~~
\bar{m}^{\mu} = {1 \over re^{\Phi / 2}} \left(0, ~ \partial U , ~
1, ~ 0 \right)
\ee
and the induced metric on $\Sigma$ is
\be
g_{z \bar{z}} = r^2 e^{\Phi} \mid_{r= U, ~ u = {\rm const.}} =
U^2 e^{\Phi} \mid_{u = {\rm const.}} ~.
\ee
There is also a complementary set of two null vectors that are both
orthogonal to those tangent vectors,
\ba
n^{\mu} & = & \left(1 \, , ~ -{F \over 2} + {\partial U ~ \bar{\partial} U
\over r^2 e^{\Phi}} \, , ~\, {\bar{\partial} U \over r^2 e^{\Phi}} \, , ~\,
{\partial U \over r^2 e^{\Phi}} \right) , \\
l^{\mu} & = & (0 \, , ~ 1 \, , ~ 0 \, , ~ 0 ) ~.
\ea
All together they form a null tetrad $(m^{\mu}, ~ \bar{m}^{\mu} ,
l^{\mu} , ~ n^{\nu} )$, which is used in the Newman-Penrose
formalism to cast the inverse Robinson-Trautman metric into the
simple form
\be
g^{\mu \nu} = m^{\mu} \bar{m}^{\nu} + \bar{m}^{\mu} m^{\nu}
- l^{\mu} n^{\nu} - n^{\mu} l^{\nu} ~.
\ee

Next, using these definitions, we compute the convergence of the two null
vectors and find, respectively,
\ba
m^{\mu} \bar{m}^{\nu} n_{(\mu ; \nu)} & = & {1 \over 2U}
\left(2 \Delta ({\rm log} U) + \Delta \Phi + {2m \over U} +
{\Lambda \over 3} U^2 \right) , \\
m^{\mu} \bar{m}^{\nu} l_{(\mu ; \nu)} & = & {1 \over r} ~.
\ea
The geodesic null congruence of Robinson-Trautman metrics is generated
by the vector field $l_{\mu} = (-1, ~ 0, ~ 0, ~0)$, which has zero shear
and twist by definition.
Setting the convergence of the null vector field $n^{\mu}$ equal to zero,
\be
2 \Delta ({\rm log} U) + \Delta \Phi + {2m \over U} +
{\Lambda \over 3} U^2 = 0 ~,
\label{apapa}
\ee
makes $\Sigma$ a marginally past trapped 2-surface and provides the
differential equation that $U(z, \bar{z})$ has to satisfy at each
instant $u = {\rm const.}$ Condition \eqn{apapa} is the generalization of
the so called Penrose-Tod equation to all values of $\Lambda$.

It turns out that $\Sigma$ is the unique
marginally trapped 2-surface on the hypersurface of constant $u$ and
it is the outer boundary of past-trapped surfaces on this hypersurface,
thus making $\Sigma$ the past apparent horizon of space-time. The
proof is outlined below in three steps, generalizing Tod's proof,
\cite{Tod}, to $\Lambda \leq 0$. The case $\Lambda > 0$ is different and it will not
be addressed here.

First, one has to establish the existence of one
{\em strictly positive solution} of the generalized Penrose-Tod equation
\eqn{apapa}. For $\Lambda = 0$, the proof of existence follows from a theorem of
Aubin, \cite{aubin}, as outlined in reference \cite{Tod}. For $\Lambda < 0$, the proof of
existence is based on the so called sub- and super-solution method adapted to Riemannian
manifolds. It is outlined in the more recent papers \cite{chech} that construct the
past apparent horizon as we do here.
Second, we note that the uniqueness of the construction follows immediately from
equation \eqn{apapa} provided that at least a {\em strictly positive solution} of this
differential equation exists, as it does. Indeed, if $U_1$ and $U_2$ are two such
distinct solutions, their ratio $R = U_1 / U_2$ will satisfy the equation
\be
2 \Delta ({\rm log} R) + {2m \over U_1} (1 - R) -
{\Lambda \over 3} U_2^2 (1-R^2) = 0 ~.
\ee
Multiplying by $1-R$ and integrating over the sphere with metric
$ds_2^2 = 2({\rm exp} \Phi) dz d \bar{z}$, one obtains the relation
\be
2 \int_{S^2} (1-R) \Delta ({\rm log} R) + 2m \int_{S^2}
{(1 - R)^2 \over U_1} +
\left( - {\Lambda \over 3} \right) \int_{S^2} U_2^2 (1+R)
(1-R)^2 = 0 ~.
\ee
The first term is non-negative as can be easily seen after integration by
parts,
\be
2 \int_{S^2} (1-R) \Delta ({\rm log} R) = \int_{S^2}
{\mid \nabla R \mid^2 \over R} ~.
\ee
The other two terms are also non-negative when $- \Lambda$ is non-negative,
since $U_1$ and $U_2$ are assumed to be strictly positive. Thus, we
necessarily have $R=1$, i.e., $U_1 = U_2$, establishing the uniqueness of
the solution when $\Lambda \leq 0$. Clearly, this argument does not apply
when $\Lambda > 0$.
Finally, as in Tod \cite{Tod}, we may express any other 2-surface $\Sigma_{(\lambda)}$
on the hypersurface of constant $u$ as $r = \lambda U$, where $\lambda$ is a
strictly positive function. The convergence of the null vector
$n_{(\lambda)}^{\mu}$ associated to the 2-surface $\Sigma_{(\lambda)}$ turns
out to be
\be
m_{(\lambda)}^{\mu} \bar{m}_{(\lambda)}^{\nu} n_{(\mu ; \nu)}^{(\lambda)} =
{1 \over 2 \lambda U} \Big[2 \Delta
\lambda + {2m \over \lambda U} (1 - \lambda) + \left(-{\Lambda \over 3}
\right) U^2 (1-\lambda^2) \Big] .
\ee
All three terms on the right hand side are negative for $\Lambda \leq 0$ when
$\lambda > 1$, i.e., when $\Sigma_{(\lambda)}$ extends to larger values of
$r$ than $\Sigma$ itself. Then, it follows that
$\Sigma$ is the outermost boundary of past-trapped surfaces on the hypersurface
of constant $u$, completing the proof.

The line element of $\Sigma$ takes the following form
\be
ds^2 (\Sigma) = \left({U \over \sigma} \right)^2 \left(d \theta^2
+ {\rm sin}^2 \theta d \phi^2 \right) ~,
\label{thelinelar}
\ee
where $\sigma$ is given by $\Phi = \Phi_0 - 2 \, {\rm log} \sigma$. The area
of $\Sigma$ is determined by taking the integral over the unit round sphere,
\be
{\rm Area}(\Sigma) = \int_{S^2} d \mu_0 \left({U \over \sigma} \right)^2 ~,
\ee
where $d \mu_0 = {\rm sin} \theta d \theta  d \phi$.  Clearly, ${\rm Area}(\Sigma)$ depends on $u$.

Using the late time expansion of solutions to the Robinson-Trautman
equation, one can determine the late time expansion
of the function $U$, and, hence, of the area of $\Sigma$. Using the first few terms of the late
time expansion of axisymmetric deformations of $S^2$ characterized by
\be
\sigma (\theta; u) = 1 + {a \over 3}(3x^2 - 1) e^{-2u/m} - {a^2 \over 234} (69 x^4
- 282 x^2 + 49) e^{-4u/m} + \cdots
\ee
with $x = {\rm cos} \theta$, we find
\ba
& & {U \over \sigma} (\theta; u) = r_{\rm h} - {a m r_{\rm h} \over 2r_{\rm h} + 3m} (3x^2 - 1) e^{-2u/m}
+ {a^2 mr_{\rm h} \over 78 (2r_{\rm h} + 3m)^2 (3r_{\rm h} + m)} \times \nonumber\\
& & ~~~~~ \left((909m^2 + 2967 m r_{\rm h} + 716 r_{\rm h}^2) x^4 - 6 \, {657m^3 + 2601 m^2 r_{\rm h} +
2046 mr_{\rm h}^2 + 472 r_{\rm h}^3 \over 2r_{\rm h} + 3m} x^2  \right. \nonumber \\
& & \left. ~~~~~ +  {2025m^4 + 7470m^3 r_{\rm h} + 4275 m^2  r_{\rm h}^2 + 142 mr_{\rm h}^3 - 408 r_{\rm h}^4
\over (2r_{\rm h} + 3m)(3m-r_{\rm h})} \right) e^{-4u/m} +  \cdots .
\label{usexp}
\ea
The result is complicated and becomes even more so to order ${\cal O} (e^{-6u/m})$.
However, the final result for the area takes very simple form
\be
{\rm Area}(\Sigma) = 4 \pi r_{\rm h}^2 \Big[1 + {16 a^2 m r_{\rm h}^2 \over
15 (3m - r_{\rm h})(2r_{\rm h} + 3m)^2} e^{-4u / m} + {\cal O}
\left(e^{-6u/m} \right) \Big] ~,
\ee
providing the leading behavior of its $u$-dependence at very late retarded times
for all $\Lambda \leq 0$. ${\rm Area}(\Sigma)$ receives no corrections at the
linearized level, since the integral of the Legendre
polynomials vanishes. Also, the leading contribution of non-linear effects comes
with positive sign which indicates that the area decreases at later retarded times, irrespective of
the size of the black hole (recall that in all cases $r_{\rm h}$ cannot exceed $2m$).

An interesting open question is whether the area of the past apparent horizon
decreases monotonically in $u$ to all orders in the expansion, thus providing a new
entropy functional for the Calabi flow on $S^2$.  Some results already exist in this direction for
$\Lambda = 0$, \cite{chow}, and we expect them to generalize to other values of
$\Lambda$. We hope to return to this problem elsewhere with more details.

\subsection{Penrose inequality}

A natural quantity in Robinson-Trautman space-times is provided by the so called Bondi
mass, \cite{BondiMass}, which is defined as follows
\be
{\cal M}_{\rm Bondi} = {m \over 4 \pi} \int_{S^2} d \mu_0 {1 \over \sigma^3} ~,
\ee
taking the integral with respect to the unit round metric on $S^2$. It incorporates the effect of gravitational
radiation in space-time.

The Bondi mass is a function of the retarded time $u$, enjoying two important
properties. First,
${\cal M}_{\rm Bondi} \geq m$, which follows immediately from Holder's inequality
\be
\left(\int_{S^2} d \mu_0 {1 \over \sigma^3} \right)^{2/3}
\left(\int_{S^2} d \mu_0 \right)^{1/3}
\geq \int_{S^2} d \mu_0 {1 \over \sigma^2} ~,
\ee
using also the fact that the evolution of the metric
\be
ds^2 (S) = {1 \over \sigma^2} \left(d\theta^2 + {\rm sin}^2 \theta
d\phi^2 \right) ~,
\ee
preserves the volume of $S^2$, i.e.,
\be
\int_{S^2} d \mu_0 {1 \over \sigma^2} = 4 \pi ~.
\ee
Second, using the Robinson-Trautman equation it follows that \cite{piotr1}
\be
{d \over du} {\cal M}_{\rm Bondi} \leq 0 ~.
\label{monot}
\ee
${\cal M}_{\rm Bondi}$ decreases monotonically along the flow,
reaching its minimal value $m$ at $u = \infty$.

Using the late time expansion of the solutions to the Robinson-Trautman equation,
as before, we find the following leading behavior of ${\cal M}_{\rm Bondi}$,
\be
{\cal M}_{\rm Bondi} = m \Big[ 1 + {2 a^2 \over 15} e^{-4u/m} + {\cal O}\left(e^{-6u/m}
\right) \Big] ~.
\ee
The Bondi mass equals $m$ at the linearized level, since the integral of the Legendre
polynomials vanishes, as before. Also, the leading contribution of non-linear effects comes
with positive sign so that ${\cal M}_{\rm Bondi}$ decreases at later retarded times,
in agreement with the general relation \eqn{monot}.

The Bondi mass is subsequently used to formulate and test the validity of Penrose inequality \cite{roger} in
Robinson-Trautman space-time. In fact, we will prove the following generalized
version of Penrose inequality in the presence of cosmological constant $\Lambda \leq 0$,
as first stated by Gibbons \cite{garry} (but see also reference \cite{bray} for an overview of the subject),
\be
16 \pi {\cal M}^2 \geq {\rm Area}(\Sigma) \left(1 - {\Lambda \over 3}
{{\rm Area}(\Sigma) \over 4 \pi} \right)^2 ~,
\label{finpen}
\ee
using the area of the past apparent horizon $\Sigma$ and letting
\be
{\cal M} = {\cal M}_{\rm Bondi} \, .
\ee
We admit, however,
that we do not yet have complete justification for choosing the same ${\cal M}_{\rm Bondi}$
for all $\Lambda \leq 0$ other than it works for Robinson-Trautman space-times.

It is convenient for our purpose to rewrite the differential equation for $\Sigma$
in the form
\be
{2m \over \sigma^3} = {U \over \sigma} \Big[1 - 2 \Delta_0
\left({\rm log} {U \over \sigma} \right) \Big] - {\Lambda \over 3}
\left({U \over \sigma} \right)^3 ~,
\ee
where $\Delta_0 = e^{-\Phi_0} \partial \bar{\partial}$ provides the Laplacian
on the unit round sphere.
Upon integration over the unit sphere with measure $d \mu_0$, we obtain
\be
8 \pi {\cal M}_{\rm Bondi} = \int_{S^2} d \mu_0 ~ {U \over \sigma}
\Big[1 - 2 \Delta_0 \left({\rm log} {U \over \sigma} \right) \Big]
- {\Lambda \over 3} \int_{S^2} d \mu_0 \left({U \over \sigma} \right)^3 ~.
\label{pen}
\ee
Since for a function $f$ on the unit sphere we have the relation
$2f \Delta_0 ({\rm log} f) = 2\Delta_0 f - |\nabla f|^2/f$, we arrive to the
following identity
\be
\int_{S^2} d \mu_0 \left(f + {|\nabla f|^2 \over f} \right)  =
 \int_{S^2} d \mu_0 ~ f \left(1 - 2 \Delta_0 ({\rm log} f) \right) ~,
\ee
dropping the contribution of $\Delta_0 f$ to the integral (it is a total derivative term on
the round $S^2$). Furthermore, as was shown by Tod, \cite{paul1}, using a combination of Sobolev
and Holder inequalities, one has the following relation for all functions $f$, provided that
they are nowhere vanishing,
\be
\int_{S^2} d \mu_0 \left(f + {|\nabla f|^2 \over f} \right)
\geq \left(4 \pi  \int_{S^2} d \mu_0 ~  f^2 \right)^{1/2} .
\ee

To prove Penrose inequality for Robinson-Trautman space-times, it suffices to choose
\be
f = {U \over \sigma}
\ee
so that the first integral on the right hand-side of equation \eqn{pen} is bounded
from below by the square root of the area functional, as
\be
\int_{S^2} d \mu_0 ~ {U \over \sigma}
\Big[1 - 2 \Delta_0 \left({\rm log} {U \over \sigma} \right) \Big] \geq
\left(4 \pi  \int_{S^2} d \mu_0 ~  \left({U \over \sigma}\right)^2 \right)^{1/2}
\equiv \sqrt{4 \pi {\rm Area}(\Sigma)} ~.
\label{step1}
\ee
The second integral on the right hand-side of equation \eqn{pen} is also bounded
from below, as can be easily seen using Holder's inequality
\be
\left(\int_{S^2} d \mu_0 \left({U \over \sigma}\right)^3  \right)^{2/3}
\left(\int_{S^2} d \mu_0 \right)^{1/3}
\geq \int_{S^2} d \mu_0 \left({U \over \sigma}\right)^2 ~.
\ee
Then, for $\Lambda < 0$ we have the inequality
\be
- {\Lambda \over 3} \int_{S^2} d \mu_0 \left({U \over \sigma} \right)^3 \geq
-{\Lambda \over 3} \, 4 \pi \left(\sqrt{{{\rm Area}(\Sigma) \over 4 \pi}} \right)^3 ~,
\ee
which when combined with \eqn{step1} yields the Penrose inequality in the form \eqn{finpen} with
${\cal M} = {\cal M}_{\rm Bondi}$.
This derivation extends the proof of Penrose inequality for Robinson-Trautman
space-times with $\Lambda = 0$, as described in reference \cite{paul1}, to the more general case
$\Lambda \leq 0$.

As consistency check, one may use the late time expansion of ${\cal M}_{\rm Bondi}$
and ${\rm Area}(\Sigma)$ obtained above to verify the validity of Penrose's inequality
for $AdS_4$ Robinson-Trautman space-times up to order ${\rm exp}(-4u/m)$. To zeroth order
we get the identity $m = m$ and to order ${\rm exp}(-4u/m)$ we get the inequality
\be
m \geq {4m r_{\rm h}^2 \over (2 r_{\rm h} + 3m)^2} ~,
\ee
which is certainly true for black holes of all sizes. Similar results hold to higher order.

\subsection{Thorne's hoop conjecture}

Another mass inequality in space-times containing a black hole is provided by the so called hoop
conjecture due to Thorne, \cite{thorne}, stating that
\be
4 \pi {\cal M} \geq C
\ee
for appropriately defined mass ${\cal M}$ and circumference $C$. We expect that the hoop conjecture
admits a suitable generalization in the presence of negative cosmological constant. We
propose that for $AdS_4$ Robinson-Trautman space-times the following inequality holds
\be
4 \pi {\cal M}_{\rm Bondi} \geq C(\Sigma) \left(1 - {\Lambda \over 3}
\left({C(\Sigma) \over 2 \pi} \right)^2  \right) ~,
\label{hoopy}
\ee
where $C(\Sigma)$ is the length of a hoop around the past apparent
horizon $\Sigma$. For definiteness, one may adopt Gibbons' proposal \cite{gibbons} for the optimal hoop
(but see also reference \cite{PaulTod} for some important related work),
choosing
\be
C(\Sigma) = \beta (\Sigma)
\ee
given by the so called Birkhoff's invariant \cite{birkhoff} of $\Sigma$. Recall that
$\beta (\Sigma)$ is a geometric invariant providing the minimum length of a hoop
(closed inextensible string)
one can slip over a spherical two-dimensional surface $\Sigma$. There always
exists a closed geodesic on $\Sigma$ - though not necessarily the shortest -
with length equal to $\beta (\Sigma)$. However, $\beta(\Sigma)$ encodes the notion of
``least circumference" in all directions.

In general, it is not possible to construct $\beta (\Sigma)$ analytically. Instead, we begin by considering
the length of the equatorial and meridional (polar) geodesic curves, using for simplicity axisymmetric
metrics on $\Sigma$. The line element \eqn{thelinelar} yields, respectively,
\be
l_{\rm e} (\Sigma)= 2 \pi \, {U \over \sigma} (\theta = \pi / 2) ~, ~~~~~
l_{\rm p} (\Sigma) = 2 \int_0^{\pi} d \theta ~ {U \over \sigma} (\theta) ~,
\label{meriando}
\ee
where the factor of $2 \pi$ in $l_{\rm e} (\Sigma)$ arises from integration over the angle $\phi$. We
further set
\be
C(\Sigma) = {\rm min} (l_{\rm e} (\Sigma), ~ l_{\rm p} (\Sigma)) ~.
\label{simpmind}
\ee

Using the late time behavior of the solutions of the Robinson-Trautman
equation, in particular the expansion \eqn{usexp} for $U/\sigma$, we find the following result
as $u \rightarrow \infty$,
\be
l_{\rm e} (\Sigma) = 2 \pi r_{\rm h} \Big[1 + {a m \over 2r_{\rm h} + 3m}
e^{-2u/m} + {\cal O} \left(e^{-4u/m} \right) \Big]
\ee
and
\be
l_{\rm p} (\Sigma) = 2 \pi r_{\rm h} \Big[1 - {am \over 2(2r_{\rm h} + 3m)}
e^{-2u/m} + {\cal O} \left(e^{-4u/m} \right) \Big] ~.
\ee
The leading order corrections to $l_{\rm e}$ and  $l_{\rm p}$ are
provided by the linear quadrupole radiation terms. The coefficient $a$ is an
arbitrary constant of either sign that determines the size of $l_{\rm e}$
relative to $l_{\rm p}$. In either case, we have to leading order,
\be
C(\Sigma) = {\rm min} (l_{\rm e} (\Sigma), ~ l_{\rm p} (\Sigma)) <
2 \pi r_{\rm h} ~.
\ee
At the same time, as noted earlier, ${\cal M}_{\rm Bondi}$ does not receive any corrections to
order ${\rm exp}(-2u/m)$ and therefore inequality \eqn{hoopy} is obviously valid. Of course,
there are corrections to $l_{\rm e} (\Sigma)$ and $l_{\rm p} (\Sigma)$ appearing to
order ${\rm exp}(-4u/m)$, as in the Bondi mass, but their effect is subdominant and cannot
change the result. Note that $C(\Sigma)$ increases with $u$ at late times.

For higher poles of gravitational radiation, the spherical part of the  Robinson-Trautman metric
assumes the form \eqn{defo},
\be
ds_2^2 = [1 + \epsilon_l (0) P_l({\rm cos} \theta) e^{-i \omega_{\rm s} u}] (d\theta^2 + {\rm sin}^2 \theta
d \phi^2) ~,
\ee
to linear approximation. Such terms become relevant when all other subleading poles of radiation are
absent from the solution. The corresponding conformal factor of the past apparent horizon has
\be
{U \over \sigma} = r_{\rm h} \left(1 + {3 \epsilon_l (0) m \over (l-1)(l+2) r_{\rm h} + 6m} \,
P_l({\rm cos} \theta) e^{-i \omega_{\rm s} u} \right)
\ee
for all $l \geq 2$ (we set $\epsilon_2(0) = -4a/3$ to compare with the previous discussion).
We can still use \eqn{simpmind} to define a hoop $C(\Sigma)$, as for the quadrupole radiation.
The equatorial and meridional closed geodesic curves, $l_{\rm e}(\Sigma)$ and $l_{\rm p}(\Sigma)$, are still
given by equation \eqn{meriando}. The integral of $P_l ({\rm cos} \theta)$ showing up in $l_{\rm p}(\Sigma)$
is always positive. Note, however, that the sign of $P_l(0)$ appearing in $l_{\rm e}(\Sigma)$ depends upon $l$:
it is negative for $l = 4k - 2$, positive for $l=4k$ and it vanishes for odd $l$. Consequently, $C(\Sigma)$
is definitely less than $2\pi r_{\rm h}$ for $l=4k-2$, as for $l=2$, and less or equal to $2\pi r_{\rm h}$ for
odd $l$. For $l=4k$, comparison of $C(\Sigma)$ to $2\pi r_{\rm h}$ depends on the sign of the arbitrary constant
$\epsilon_l(0)$, making the definition \eqn{simpmind} of $C(\Sigma)$ inappropriate.
Then, in that case, one has to consider lassoing $\Sigma$ sideways, trying to find better ways to satisfy the
inequality \eqn{hoopy}, which to linear order reads
\be
C(\Sigma) \left(1 - {\Lambda \over 3}
\left({C(\Sigma) \over 2 \pi} \right)^2  \right) \leq 2\pi r_{\rm h} \left(1 - {\Lambda \over 3} r_{\rm h}^2 \right)
= 4 \pi m ~.
\ee
We will not pursue this calculation here, since there is a systematic way to prove the inequality
for $l=4k$, as for $l=4k-2$, at linear as well as non-linear level, as will be seen next.
The Birkhoff length need not be equal to any one of the choices described above.

For Robinson-Trautman space-times that are axisymmetric and also admit antipodal symmetry, so that only
terms with even $l$ appear in the linearized approximation, it is
possible to give a general proof of the hoop conjecture \eqn{hoopy} for appropriate choice of $C(\Sigma)$.
The proof puts in context the previous discussion and it is applicable to the full non-linear regime
of Einstein equations. In that case, $\Sigma$ also has antipodal symmetry by construction and, thus, its area
is bounded from below
\be
{\rm Area}(\Sigma) \geq {l^2(\Sigma) \over \pi}
\ee
in terms of the length $l(\Sigma)$ of the shortest non-trivial closed geodesic on $\Sigma$ (the area and
length are both computed with respect to a given metric on $\Sigma$). This is a rather old mathematical
result due to Pu, \cite{pu}, which is applicable to all metrics on 2-spheres with antipodal symmetry. Then,
the hoop inequality \eqn{hoopy} follows immediately from the generalized form of Penrose inequality
\eqn{finpen} that was proved before, choosing $C(\Sigma) = l(\Sigma)$. As such, it generalizes the relation between
Penrose and Thorne inequalities described in \cite{gibbons}, but now in the presence of cosmological constant
$\Lambda \leq 0$. Gibbon's version of the hoop conjecture
provides a more stringent lower bound, since the Birkhoff invariant is not necessarily the length of the
shortest closed geodesic, i.e., $\beta (\Sigma) \geq l(\Sigma)$, but we do not yet have a general proof of
it for Robinson-Trautman space-times.

A related interesting open question is whether the length of the hoop, being $l(\Sigma)$ or $\beta(\Sigma)$,
exhibits monotonicity under Calabi flow that governs the evolution of the past apparent horizon.
%Likewise, one may ask the same question for the Birkhoff length
%and the length of the shortest closed geodesic defined on the spherical part of the Robinson-Trautman metric
%undergoing area preserving deformations.

\section{Holographic aspects}
\setcounter{equation}{0}

In this section, we briefly review the construction of the energy-momentum
tensor for AdS gravity, based on holography
\cite{malda, igor, ed1, Henningson:1998gx, kraus, skenderis1, skenderis2, skenderis3, Papadimitriou:2005ii},
and apply it to the general class of Robinson-Trautman metrics with negative cosmological constant. Then,
the resulting energy-momentum tensor is specialized to the Schwarzschild solution and its perturbations with respect to
the algebraically special modes and determine their effective viscosity. The complete expression also includes non-linear
effects of gravitational radiation, which can be described systematically at late times.

\subsection{Energy-momentum tensor}

The $AdS_4$ Robinson-Trautman solution is an example of an asymptotically locally AdS space-time\footnote{We
refer the reader to section 3 of the original work \cite{skenderis3} for the precise definition. Note, however, that asymptotically
locally AdS space-times were called asymptotically AdS in that reference. Here, as in most of the literature,
we reserve the terminology asymptotically AdS only for space-times that look like AdS close to the boundary.}. For these
space-times the metric near the conformal boundary $\mathscr{I}$ takes the Fefferman-Graham form
\be \label{FG}
ds^2 = -\frac{3}{\Lambda} \Big[\frac{d\varrho^2}{\varrho^2} +  \frac{1}{\varrho^2} \left( g_{(0)ab}(x)
+ \varrho^2 g_{(2)ab}(x) + \varrho^3 g_{(3)ab}(x)+ \cdots \right) dx^a dx^b\Big] ~,
\ee
where
$\sqrt{-3/\Lambda}$ is the AdS radius and $\varrho=0$ is the location of the conformal boundary. If the conformal boundary
is topologically $\mathbb{R} \times S^2$ and $g_{(0)}$ is conformally flat then the space-time is asymptotically AdS, otherwise
it is only asymptotically locally AdS. Asymptotically locally AdS space-times come equipped with a conserved symmetric tensor,
$T_{ab}$, which in even dimensions is also traceless, i.e.,
\be
\nabla^{b} T_{ab}=0, \qquad T^a_a =0 ~,
\ee
where $\nabla^a$ is the covariant derivative associated to $g_{(0)ab}$. This tensor can be extracted from the asymptotics of
the solution, \cite{skenderis1},
\be \label{tij}
T_{ab} = - \frac{3}{2 \kappa^2} \left(-\frac{3}{\Lambda}\right) g_{(3)ab}
\ee
and is called the holographic energy-momentum tensor because it also represents the expectation value of the
energy-momentum tensor in the dual QFT, \cite{Henningson:1998gx, kraus, skenderis1, skenderis3}, as
\be
\langle T_{ab} \rangle = {2 \over \sqrt{-{\rm det} g_{(0)}}} {\delta S_{\rm ren} \over
\delta g_{(0)}^{ab}} ~.
\ee
$S_{\rm ren}$ is the on-shell gravitational action supplemented by covariant counter-terms to remove the infinite volume
divergences, \cite{Henningson:1998gx,kraus, skenderis1}. It provides a well defined prescription for implementing the
Brown-York proposal for quasi-local energy, \cite{york}, without using a reference space-time.
One can show that if the space-time possesses asymptotic Killing vectors, there will be conserved charges, which one
can compute using $T_{ab}$, \cite{Papadimitriou:2005ii} (but see also references \cite{kraus, skenderis2}). Specializing to
asymptotically AdS space-times, which is the only case that was discussed in all generality in earlier works, one recovers
previous prescriptions for the computation of conserved charges.

Thus, in order to compute the holographic energy-momentum tensor, it suffices to change to Fefferman-Graham coordinates
near the conformal boundary and then extract the coefficient $g_{(3)}$. The Fefferman-Graham coordinates are Gaussian normal
coordinates centered at the conformal boundary. To reach these coordinates we need to ensure that
$g_{\varrho \varrho}=1/\varrho^2$ and the off-diagonal terms between $\varrho$ and $x^a$ are zero up to sufficiently high order.
In practice, we first change variables $u = t-r_{\star}$, introducing explicit $r$-dependence into the function
$\Phi (z, \bar{z}; u)$ so that the evolution is taken with respect to the real time $t$ rather than the retarded time $u$;
of course, we have $\partial_u \Phi = \partial_t \Phi$. Since the conformal boundary is located at $r_{\star}=0$,
we are led to perform the following change of variables,
\begin{eqnarray}
r_{\star} &\to& \varrho - \frac{1}{2} (\pa_t \hat{\F}) \varrho^2 + \Big[\frac{\Lambda}{9}
+ \frac{3}{16} (\pa_t \hat{\F})^2 + \frac{1}{4} \pa^2_t \hat{\F}
+\frac{\Lambda}{12} \hat{\D} \hat{\F}\Big] \varrho^3
-\frac{1}{48}\Big[3 (\pa_t \hat{\F})^3 \Big. \\
&&+ 8 \pa_t \hat{\F} (3 + \pa_t^2 \hat{\F})
+\Big. 4 (2 m + \pa_t^3 \hat{\F})
+ 4 (3 \hat{\D} \hat{\F} \, \pa_t \hat{\F} + 2 \pa_t (\hat{\D} \hat{\F}))\Big] \varrho^4
+ \co(\varrho^5) ~, \nonumber \\
t &\to& t - \frac{1}{2} (\pa_t \hat{\F}) \varrho^2 + \frac{1}{6} \Big[
\frac{2 \Lambda}{3} + (\pa_t \hat{\F})^2 + \pa^2_t \hat{\F}
+ \frac{2 \Lambda}{3} \hat{\D} \hat{\F}\Big] \varrho^3
-\frac{1}{32}\Big[2 (\pa_t \hat{\F})^3 \Big. \\
&& + \pa_t \hat{\F} \left(\frac{16 \Lambda}{3} + 5 \pa_t^2 \hat{\F}\right)
+\Big. 2 \pa_t^3 \hat{\F}
+ \frac{2 \Lambda}{3} (4 \hat{\D} \hat{\F} \, \pa_t \hat{\F} + 3 \pa_t (\hat{\D} \hat{\F}))\Big] \varrho^4
+ \co(\varrho^5) ~, \nonumber \\
z &\to& z + e^{-\hat{\F}} \pa_{\bar{z}} \Big[
-\frac{\Lambda}{18} (\pa_t \hat{\F}) \varrho^3
+ \frac{\Lambda}{48} \left(\pa_t^2 \F
-\frac{1}{4} (\pa_t \hat{\F})^2 + \frac{\Lambda}{3} \hat{\D} \hat{\F}\right) \varrho^4\Big]  +
\co(\varrho^5) ~, \\
\bar{z} &\to& \bar{z} + e^{-\hat{\F}} \pa_{z} \Big[
-\frac{\Lambda}{18} (\pa_t \hat{\F}) \varrho^3
+ \frac{\Lambda}{48} \left(\pa_t^2 \F
-\frac{1}{4} (\pa_t \hat{\F})^2 + \frac{\Lambda}{3} \hat{\D} \hat{\F}\right) \varrho^4\Big]
+ \co(\varrho^5) ~,
\end{eqnarray}
where $\hat{\F}$ is the boundary value of $\F$,
\be
\hat{\F}(z,\bar{z};t) = \lim_{r_{\star} \to 0} \F(z,\bar{z};u).
\ee
Note that $\hat{\Phi}$ has the same functional form as $\Phi$, since $u = t-r_{\star}$,
and it satisfies the fourth-order diffusion equation in real time,
\be
3m \partial_t \hat{\Phi} + \hat{\Delta} \hat{\Delta} \hat{\Phi} = 0 ~,
\ee
where $\hat{\Delta}$ is the Laplacian on the spherical
spatial slices of the three-dimensional boundary $\mathscr{I}$,
\be
\hat{\Delta} = e^{- \hat{\Phi}} \partial_z \partial_{\bar{z}}~.
\ee

After this change of variables, the space-time metric takes the form (\ref{FG})
with the metric components $g_{(0)}, g_{(2)}$ and $g_{(3)}$ given by
\begin{eqnarray} \label{asymp}
ds_{(0)}^2 &=& - dt^2 -\frac{6}{\Lambda} e^{\hat{\F}} dz d\bar{z}\\
ds_{(2)}^2 &=& {1 \over 2} \Big[\frac{1}{4} (\pa_t \hat{\F})^2 + \pa_t^2 \F
- \frac{\Lambda}{3} \hat{\D} \hat{\F}\Big] dt^2
+ \Big[\frac{3}{4\Lambda} (\pa_t \hat{\F})^2 + \hat{\D} \hat{\F}\Big]
e^{\hat{\F}} dz d\bar{z}
+ [(\pa_z \pa_t \hat{\F}) dt dz]_{\rm c.c.} \nonumber\\
ds_{(3)}^2 &=&  -\frac{4 m \Lambda^2}{27}dt^2
+ \frac{4 m \Lambda}{9} e^{\hat{\F}} dz d\bar{z}
-{1 \over 3} \Big[\frac{2 \Lambda}{3} \pa_z (\hat{\D} \hat{\F}) dt dz
+ \pa_t \left(\frac{1}{2}
(\pa_z \hat{\F})^2 - \pa_z^2 \hat{\F} \right) dz^2 \Big]_{\rm c.c.} \nonumber
\end{eqnarray}
where c.c. means that one has to add the complex conjugate to the corresponding bracketed terms.

The metric $g_{(0)}$ is the representative of the boundary conformal structure. Thus, we set
\be
ds_{\mathscr{I}}^2 = - dt^2 -\frac{6}{\Lambda} e^{\hat{\F}} dz d\bar{z} ~.
\ee
To check whether this metric is conformally flat, we compute its Cotton tensor.
Recall that the Cotton tensor of a three-dimensional metric $\gamma_{ab}$ is defined as follows,
\ba
C^{ab} & = & {1 \over 2 \sqrt{- {\rm det} \gamma}} \left( \epsilon^{acd}
\nabla_c {R^b}_d + \epsilon^{bcd} \nabla_c {R^a}_d \right) \nonumber\\
& = & {\epsilon^{acd} \over \sqrt{- {\rm det} \gamma}} \nabla_c
\left({R^b}_d - {1 \over 4} {\delta^b}_d R \right) ,
\label{cotta}
\ea
letting $\epsilon^{t z \bar{z}} = i$ (equivalently, $\epsilon^{t \theta \phi} = 1$
in spherical coordinates). It is a symmetric and traceless tensor that is
covariantly conserved identically, without employing the classical equations
of motion. The density $\sqrt{{\rm det} \gamma } ~ {C^a}_b$ remains invariant under local conformal
changes of the metric $\gamma_{ab}$ and it vanishes if and only if the metric is
conformally flat. In our case, it takes the form
\be
C_{zz} = - {i \over 4} \partial_t
\left((\partial_z \hat{\Phi})^2 - 2 \partial_z^2
\hat{\Phi} \right) , ~~~~~
C_{tz} = -i {\Lambda \over 6} \partial_z (\hat{\Delta}
\hat{\Phi})
\ee
and
\be
C_{\bar{z} \bar{z}} = - \bar{C}_{zz} ~, ~~~~~
C_{t \bar{z}} = - \bar{C}_{tz} ~,
\ee
whereas all other components vanish identically. Thus, $g_{(0)}$ is not conformally flat for general
Robinson-Trautman solutions; these solutions are indeed only asymptotically locally AdS.

The asymptotic analysis of reference \cite{skenderis1} implies that the
$g_{(2)}$ should be given by
\be \label{g2}
g_{(2)ab} = - {\cal R}_{ab} + \frac{1}{4} {\cal R} g_{(0)ab} \, ,
\ee
where ${\cal R}_{ij}, {\cal R}$ are the Ricci and scalar curvatures of $g_{(0)}$.
Indeed, using
\begin{eqnarray}
&&{\cal R}_{tt} = -\pa_t^2 \hat{\F} - \frac{1}{2} (\pa_t \hat{\F})^2, \quad
{\cal R}_{z\bar{z}}= -\frac{1}{2} e^{\hat{\F}} \left( \frac{3}{\Lambda} \left(
\pa_t^2 \hat{\F} + (\pa_t \hat{\F})^2\right)+ 2 \hat{\D} \hat{\F}\right),
\nonumber \\
&& {\cal R}_{tz} = -\frac{1}{2} \pa_t \pa_z \hat{\F} \, , \quad
{\cal R}_{t\bar{z}} = -\frac{1}{2} \pa_t \pa_{\bar{z}} \hat{\F} \, , \quad
{\cal R} = 2 \pa_t^2 \hat{\F} + \frac{3}{2} (\pa_t \hat{\F})^2+ \frac{2 \Lambda}{3} \hat{\D} \hat{\F} \, ,
\end{eqnarray}
one can check that this is the case.

Finally, the energy-momentum tensor follows from $g_{(3)}$ via equation (\ref{tij}).
We obtain
\ba
& & \kappa^2 T_{tt} = -{2m \Lambda \over 3} ~, ~~~~~~
\kappa^2 T_{tz} = -{1 \over 2} \partial_z (\hat{\Delta}
\hat{\Phi})  \, , \\
& & \kappa^2 T_{z \bar{z}} = m e^{\hat{\Phi}} , ~~~~~~
\kappa^2 T_{zz} = - {3 \over 4 \Lambda} \partial_t
\left((\partial_z \hat{\Phi})^2 - 2 \partial_z^2
\hat{\Phi} \right) ,
\ea
whereas
\be
T_{t \bar{z}} = \bar{T}_{tz}  ~, ~~~~~~
T_{\bar{z} \bar{z}}  = \bar{T}_{zz}~.
\ee
Note that the energy-momentum tensor is traceless and conserved
using the classical equations of motion, ${T^a}_a = 0$ and $\nabla^a T_{ab} = 0$,
as required on general grounds.

Note also the following relations between the components of the energy-momentum and the Cotton tensor,
\ba
& & C_{zz} = i {\Lambda \over 3} \kappa^2 T_{zz} ~, ~~~~~
C_{\bar{z} \bar{z}} = - i {\Lambda \over 3} \kappa^2 T_{\bar{z} \bar{z}}
~, \\
& & C_{tz} = i {\Lambda \over 3} \kappa^2 T_{tz} ~, ~~~~~
C_{t \bar{z}} = - i {\Lambda \over 3} \kappa^2 T_{t \bar{z}} ~.
\ea
Despite appearances, all components of the energy-momentum as well as the Cotton tensor
are real when written in spherical coordinates $(\theta, \, \phi)$.

\subsection{Linearization of $T_{ab}$}

We first apply the formulae to the simple example of static $AdS_4$
Schwarzschild solution that serves as reference
frame to study the effect of linear (as well as non-linear) perturbations.
In this case, the three-dimensional metric on $\mathscr{I}$ describes an Einstein universe, which is
conformally flat, written in spherical coordinates as
\be
ds_{\mathscr{I}}^2 = -dt^2 -{3 \over \Lambda}(d\theta^2 + {\rm sin}^2
\theta d\phi^2) ~.
\ee
The renormalized energy-momentum tensor  has the following non-vanishing components
\be
\kappa^2 T_{tt}^{(0)} = -{2 m \Lambda \over 3} ~, ~~~~~
\kappa^2 T_{\theta \theta}^{(0)} = m ~, ~~~~~
\kappa^2 T_{\phi \phi}^{(0)} = m ~ {\rm sin}^2 \theta ~,
\ee
reproducing the expressions already known in the literature.
The superscript $(0)$ is used for reference to the static background.

Next, we consider the algebraically special modes with purely imaginary frequencies
\be
\omega_{\rm s} = -i
{(l-1)l(l+1)(l+2) \over 12m}
\ee
that arise to linear order as perturbations of $AdS_4$ Schwarzschild metric and compute their contribution to
the energy-momentum tensor. Assuming, for simplicity, that the perturbations are axially symmetric, so that
\be
\hat{\Phi} (\theta; t) = e^{-i \omega_{\rm s} t} P_l ({\rm cos \theta})
+ 4 {\rm log}\left({\rm sin} {\theta \over 2} \right) ,
\ee
the three-dimensional metric on $\mathscr{I}$ takes the form
\be
ds_{\mathscr{I}}^2 = -dt^2 - {3 \over \Lambda} [1 + e^{-i \omega_{\rm s} t}
P_l({\rm cos} \theta)] (d\theta^2 + {\rm sin}^2 \theta d\phi^2) ~,
\ee
whereas $T_{ab}$ is
\be
T_{ab} = T_{ab}^{(0)} + \delta T_{ab} ~.
\ee
We have $\delta T_{t \phi} = 0 = \delta T_{\theta \phi}$, as consequence of
axial symmetry, and $\delta T_{tt} = 0$. The remaining components turn out to be
\ba
\kappa^2 \delta T_{\theta \theta} & = &
m \left(1 + {3i \omega_{\rm s} \over 4m \Lambda} l (l+1) \right)
e^{-i \omega_{\rm s} t} P_l ({\rm cos} \theta) + \nonumber\\
& & {3i \omega_{\rm s} \over 2 \Lambda} e^{-i \omega_{\rm s} t} {\rm cot} \theta ~
\partial_{\theta} P_l ({\rm cos} \theta) \\
\kappa^2 \delta T_{\phi \phi} & = &
m \left(1 - {3i \omega_{\rm s} \over 4m \Lambda} l (l+1) \right)
e^{-i \omega_{\rm s} t} {\rm sin}^2 \theta ~ P_l ({\rm cos} \theta) - \nonumber\\
& & {3i \omega_{\rm s} \over 2 \Lambda} e^{-i \omega_{\rm s} t} {\rm sin} \theta
{\rm cos} \theta ~ \partial_{\theta} P_l ({\rm cos} \theta) \\
\kappa^2 \delta T_{t \theta} & = & {1 \over 4} (l-1)(l+2)
e^{-i \omega_{\rm s} t} \partial_{\theta} P_l ({\rm cos} \theta) ~.
\ea

\subsection{Effective viscosity}

We are now in position to examine whether the algebraically special models admit an effective
hydrodynamic representation.

Given a conserved energy-momentum tensor with non-negative energy density,
i.e, assuming $T_{ab} u^a u^b \geq 0$ for all time-like vectors, it is always
possible to solve the eigenvalue problem
\be
T_{ab} u^b = - \rho u_a
\ee
in terms of a unique time-like vector $u^a$ that is normalized as $u^a u_a = -1$.
This vector defines a frame, usually called the energy or Landau frame, which we
adopt in the following. The energy-momentum tensor of a fluid admits the decomposition
\be
T^{ab} = \rho u^a u^b + p \Delta^{ab} + \Pi^{ab} ~,
\ee
where
\be
\Delta^{ab} = u^a u^b + g^{ab}
\ee
and $\rho$, $p$ are the energy density and pressure, respectively, in the
local rest frame. For conformal fluids in three space-time dimensions $\rho$ and $p$
obey the equation of state $\rho = 2 p$.
$\Pi^{ab}$ is a transverse tensor, $u_a \Pi^{ab} = 0$, that describes the viscous
part of the energy-momentum tensor of a fluid and it admits an expansion
in derivatives of $u^a$,
\be
\Pi^{ab} = \Pi_{(1)}^{ab} + \Pi_{(2)}^{ab} + \cdots ~.
\ee

First order hydrodynamics is concerned with the structure of $\Pi_{(1)}^{ab}$ and
is well studied (see, for instance, the classic reference \cite{landau}).
For notational purposes, we use the bracketed tensor associated
to any second rank tensor $A^{ab}$ in three dimensions,
\be
A^{<ab>} = {1 \over 2} \left(\Delta^{ac} \Delta^{bd}
(A_{cd} + A_{dc}) - \Delta^{ab} \Delta^{cd}
A_{cd} \right) ,
\ee
which is transverse, $u_a A^{<ab>} = 0$, and traceless, $g_{ab} A^{<ab>} = 0$.
Then, the viscosity tensor has the following general form in relativistic first order hydrodynamics,
\be
\Pi_{(1)}^{ab} = - \eta \sigma^{ab} - \zeta \Delta^{ab} (\nabla_c u^c) ~,
\ee
where
\be
\sigma^{ab} = 2 \nabla^{<a} u^{b>}
\ee
expresses the symmetric, transverse and traceless part of $\Pi^{ab}$
up to first derivatives in $u^a$. The coefficients $\eta$ and $\zeta$
depend, in general, on $\rho$ and they are called shear and bulk viscosity,
respectively. Conformal fluids have $\zeta = 0$,
the speed of sound in three dimensions is $1/\sqrt{2}$ and $\eta$ is
non-zero. This is precisely the case we are considering here.

The energy-momentum tensor associated to the static $AdS_4$ black hole
represents a perfect fluid with velocity vector $u_a =(-1, \, 0, \, 0)$
and energy density
\be
\kappa^2 \rho = -{2m \Lambda \over 3} ~.
\ee
Switching on perturbations, due to algebraically special modes,
yield a velocity vector in the energy frame with components
\be
u_t = -1 ~, ~~~~~ u_{\phi} = 0 ~,
\ee
as before, whereas $u_{\theta}$ changes to
\be
u_{\theta} = {1 \over 4 m \Lambda} (l-1)(l+2) e^{-i \omega_{\rm s} t}
\partial_{\theta} P_l ({\rm cos} \theta) ~.
\ee
The energy density $\rho$ is not affected by the perturbations. Then,
it is straightforward to compute the deviation from the perfect fluid
form. The corresponding viscosity tensor has components
\be \label{pi_theta}
\kappa^2 \Pi_{\theta \theta}^{(1)} = {1 \over 16 m \Lambda} (l-1)l(l+1)(l+2)
e^{-i \omega_{\rm s} t}  [l(l+1) P_l ({\rm cos} \theta) +
2 {\rm cot} \theta ~ \partial_{\theta} P_l ({\rm cos} \theta)] ~,
\ee
whereas
\be \label{pi_phi}
\Pi_{\phi \phi}^{(1)} = - {\rm sin}^2 \theta ~ \Pi_{\theta \theta}^{(1)} ~.
\ee
All other components vanish identically.

The shear viscosity associated to the $l$-th algebraically special mode
turns out to be
\be
\kappa^2 \eta = {1 \over 4} l(l+1) ~,
\ee
as can be easily verified. The dependence of $\eta$ upon $l$ implies that
the ratio of shear viscosity to the entropy density of an $AdS_4$ black hole,
\be
{\eta \over s} = {4 \over r_{\rm h}^2} \left(-{3 \over \Lambda} \right)
\eta = {l(l+1) \over 8 \pi} {r_{\rm h} \over 2m - r_{\rm h}} ~,
\ee
also depends upon $l$ and thus $\eta / s$ differs from the celebrated KSS
value $1 /4 \pi$, \cite{kss1,kss2}.
In fact, large black holes with sufficiently low $l$, so that
\be
{m \over r_{\rm h}} > {l(l+1) + 2 \over 4} ~,
\label{vkss}
\ee
appear to violate the KSS bound, since
\be
{\eta \over s} < {1 \over 4 \pi} ~.
\ee
Large black holes with higher values of $l$ satisfy $\eta / s
\geq 1 / 4 \pi$. On the other hand, small black holes satisfy the
inequality $\eta / s > 1 / 4 \pi$ for all values of $l$.

In AdS/CFT correspondence, linear perturbations around AdS black black satisfying Dirichlet boundary
conditions at the conformal boundary and incoming boundary conditions at the horizon compute retarded thermal
correlators. From those one can compute the shear viscosity via the Kubo formula. For such fluctuations there is
a holographic derivation of the KSS bound. To obtain the correct physical interpretation of the results for the
algebraically special modes one would need to understand first what exactly are these fluctuation computing in
the dual QFT.

Note also that there is some correlation between
the global structure of Robinson-Trautman space-times and the
violation of the KSS bound by the algebraically special modes.
As explained before, the algebraically special perturbations of large $AdS_4$
black holes do not have a Kruskal
extension across the horizon for sufficiently small values
of $l$. In fact, for the values of $l$ constrained by inequality
\eqn{vkss} we have $2i \omega_{\rm s} \delta_{\rm h} <1$, in the
notation of section 3.1, and this leads to divergencies on the
null hypersurface $\mathscr{H}^+$ that are correlated with the result
$\eta / s < 1/ 4 \pi$. The divergencies on $\mathscr{H}^+$ wash away
for very high values of $l$, since $2i \omega_{\rm s} \delta_{\rm h} >2$,
in which case  $\eta / s > 1 / 4 \pi$. However, there are intermediate
values of $l$ that lead to divergencies on $\mathscr{H}^+$,
though milder, but they are still capable to produce $\eta / s > 1 / 4 \pi$.

We can rewrite the result for the ratio $\eta / s$ in simple
form that is applicable to very large $AdS_4$ black holes.
It has been observed in the literature, as result of numerical investigations,
that very large $AdS_4$ black holes exhibit another set of purely dissipative modes with frequencies
\be
\Omega_{\rm s} = - i {(l-1)(l+2) \over 3 r_{\rm h}}
\label{riorio}
\ee
for all $l \geq 2$, \cite{berti, konop} (but see also reference \cite{miranda} for a derivation).
They correspond to {\em axial} perturbations, satisfying perfectly reflecting Dirichlet boundary
conditions\footnote{Obviously, the same set of modes also arise for polar
perturbations satisfying mixed boundary conditions that are supersymmetric
partner to those particular axial perturbations of very large $AdS_4$ black holes.}.
Also, they are true hydrodynamic modes of very large $AdS_4$ black holes, saturating the KSS bound
$\eta/s = 1/4 \pi$, \cite{Michalogiorgakis:2006jc}.
Then, using $\Omega_{\rm s}$, the ratio of shear viscosity to the entropy density of the algebraically special modes
$\omega_{\rm s}$ of very large (flat) black holes takes the following form
\be
{\eta \over s} = {1 \over 4 \pi} {\omega_{\rm s} \over \Omega_{\rm s}} ~.
\ee
The deviations from the KSS bound depend on the size of $\omega_{\rm s}$
relative to $\Omega_{\rm s}$ for any given $l$. Actually, as pointed out in section 2,
the values $\Omega_{\rm s}$ provide the characteristic inverse time
scale, call it $\tau_{\rm R}$, for the decay of linear perturbations of
a round sphere under the normalized Ricci flow in the multi-pole
expansion\footnote{Note, however, that unlike the Robinson-Trautman
metric, which provides a non-linear extension of the algebraically special
modes, there is no such extension known for the $\Omega_{\rm s}$ modes.
An interesting question is whether the normalized Ricci flow
can be embedded into Einstein equations with cosmological
constant $\Lambda <0$, thus providing a new class of exact radiative
metrics that settle down to very large black holes (at least up to
$1/r_{\rm h}^2$ corrections). If so, there will be an analytic explanation
for the spectrum $\Omega_{\rm s}$.}. Likewise, $\omega_{\rm s}$ provide
the characteristic inverse time scale, call it $\tau_{\rm C}$, for the
decay of linear perturbations of a round sphere under the Calabi flow,
and, therefore,
\be
{\omega_{\rm s} \over \Omega_{\rm s}} = {\tau_{\rm R} \over
\tau_{\rm C}} ~.
\label{ratiorc}
\ee
This ratio is used to compare how fast the curvature perturbations dissipate under the
two curvature flows. For large black holes it becomes less than
1 for low lying $l$.

We emphasize  that the late time asymptotic
expansion of the energy-momentum tensor is not the same as the derivative
expansion of the viscosity tensor $\Pi^{ab}$. Recall that first order
hydrodynamics captures effects to first order in the derivatives while
higher order hydrodynamics associated to $\Pi_{(n)}^{ab}$ terms captures terms
of order $n$ in the derivatives. Each of $\Pi_{(n)}^{ab}$ can be expanded in a basis of independent tensors
and the coefficients of these tensors defined the $n$-th order transport coefficients.
When we evaluate $\Pi^{ab}$ using the late time expansion of the metric higher order
$\Pi_{(n)}^{ab}$ may become degenerate with $\Pi_{(1)}^{ab}$ and the effective viscosity computed via the late time expansion would then be equal to the true viscosity plus a combinations of higher order transport coefficients. Thus, the
hydrodynamic representation of the linear gravitational perturbations at late times is not in general sufficient in order to extract the true shear viscosity of the system.

\subsection{Late time expansion of $T_{ab}$} \label{late_exp_T}

The late time expansion of solutions to Robinson-Trautman equation, as described in
section 2.4, can be used to obtain a similar expansion for the energy-momentum tensor and the
boundary metric. In principle, one can compute recursively
all higher order corrections to the energy-momentum tensor away from the
static black hole solution, going beyond the linear approximation, but the expansion
will be truncated here to order ${\rm exp}(-4m/t)$ for illustration.
The result captures the effect of the most dominant non-linear terms beyond the
linear quadrupole approximation. We have
\be
T_{ab} = T_{ab}^{(0)} - {4a \over 3} \delta T_{ab}^{(1)}
+ {8a^2 \over 39} \delta T_{ab}^{(2)} ,
\label{laertisa}
\ee
where $\delta T_{ab}^{(1)}$ is the contribution of the lowest lying algebraically special mode,
$l=2$, which is of order ${\rm exp} (-2t/m)$, and $\delta T_{ab}^{(2)}$ denotes the first
non-linear corrections of order ${\rm exp} (-4t/m)$.

Keeping $a$ arbitrary and using spherical coordinates we have the following
non-vanishing quadrupole perturbations, specializing our previous results to $l=2$,
\ba
\kappa^2 \delta T_{\theta \theta}^{(1)} & = & {m \over 2}
\left(3 {\rm cos}^2 \theta -1 - {9 \over m^2 \Lambda} {\rm sin}^2 \theta
\right) e^{-2t/m} ~, \\
\kappa^2 \delta T_{\phi \phi}^{(1)} & = & {m \over 2}
\left(3 {\rm cos}^2 \theta -1 + {9 \over m^2 \Lambda} {\rm sin}^2 \theta
\right) {\rm sin}^2 \theta ~ e^{-2t/m} ~, \\
\kappa^2 \delta T_{t \theta}^{(1)} & = & -3 {\rm sin} \theta
{\rm cos} \theta ~ e^{-2t/m} ~.
\ea
The first non-linear terms contributing to the renormalized energy-momentum tensor are
\ba
\kappa^2 \delta T_{\theta \theta}^{(2)} & = &
{m \over 2} \left(35 {\rm cos}^4 \theta - 43 {\rm cos}^2 \theta + {22 \over 3}
\right) e^{-4t/m} + \nonumber\\
& & {18 \over m \Lambda} \left(9 {\rm cos}^4 \theta - 14 {\rm cos}^2 \theta +
5 \right) e^{-4t/m} ~, \\
\kappa^2 \delta T_{\phi \phi}^{(2)} & = &
{m \over 2} \left(35 {\rm cos}^4 \theta - 43 {\rm cos}^2 \theta + {22 \over 3}
\right) {\rm sin}^2 \theta ~ e^{-4t/m} - \nonumber\\
& & {18 \over m \Lambda} \left(9 {\rm cos}^4 \theta - 14 {\rm cos}^2 \theta +
5 \right) {\rm sin}^2 \theta ~ e^{-4t/m} ~, \\
\kappa^2 \delta T_{t \theta}^{(2)} & = &
2 {\rm sin} \theta {\rm cos} \theta \left(21{\rm sin}^2 \theta + 1 \right)
~ e^{-4t/m} .
\ea
Since axial symmetry has been used in the calculation, the $(t \phi)$
and $(\theta \phi)$ components of $T_{ab}^{(2)}$ are zero. Also, $\delta T_{t t}^{(2)}$ vanishes.
As for the three-dimensional metric on $\mathscr{I}$, it takes
the following form, up to this order,
\ba \label{2nd_corr}
ds_{\mathscr{I}}^2 & = & -dt^2 - {3 \over \Lambda} [1 - {2a \over 3}
\left(3 {\rm cos}^2 \theta - 1 \right) e^{-2t/m} + \nonumber\\
& & {4a^2 \over 39} \left(35 {\rm cos}^4 \theta - 43 {\rm cos}^2 \theta +
{22 \over 3} \right) e^{-4t/m}] (d\theta^2 + {\rm sin}^2 \theta d\phi^2) ~.
\ea
Subleading non-linear corrections of order ${\rm exp}(-6m/t)$ or higher
are more cumbersome to compute explicitly.

The Calabi flow on $S^2$ incorporates all higher viscosity terms in closed
geometric form. One should note that these higher order corrections need not capture purely hydrodynamic modes.

\subsection{Entropy production}

In this subsection we discuss entropy production as the system approaches equilibrium.
As is well known \cite{landau}, a dissipative system with positive shear viscosity has an entropy current
with non-negative divergence given by
\begin{equation}
s^a = s u^a
\end{equation}
where $s$ is the local entropy density.

In the hydrodynamic regime the system is in local thermal equilibrium
and the local proper energy $\rho$, the pressure $p$, the local temperature ${\cal T}$ and the local entropy
density $s$ satisfy the standard thermodynamic relations,
\be \label{th_eq}
\rho + p = {\cal T} s, \qquad d \rho = {\cal T} ds, \qquad d p = s \, d {\cal T} ~.
\ee
Since we have a conformal fluid, the thermodynamic relations easily integrate to yield 
\be
s = \gamma {\cal T}^2 , ~~~~~~ \rho = 2p = {2 \gamma \over 3} \, {
\cal T}^3 . 
\ee 
The bulk solutions need to approach
large AdS black holes at late times, for which $r_h \gg L = \sqrt{-3/\Lambda}$, in order to be in the hydrodynamic
regime \cite{Bhattacharyya:2007vs}. In this limit, the equilibrium values of $\rho$, $p$, ${\cal T}$ and $s$ are
\be \label{eq_values}
\rho_{0} = \frac{1}{\kappa^2} r_h^3 \left(-{\Lambda \over 3}\right)^2 = 2 p_{0}, \quad
{\cal T}_{0} = \frac{3}{4 \pi} r_h  \left(-{\Lambda \over 3}\right) \quad
s_{0} = {1 \over 4} r_h^2  \left(-{\Lambda \over 3}\right),
\ee
fixing $\gamma = - 4\pi^2/3\Lambda$. In this limit, we also have
$r_{\rm h}^3 = -6m/ \Lambda$, as discussed in Appendix B.
To work out the leading order corrections to the equilibrium values (\ref{eq_values}), we set
\be
\rho=\rho_{0}+\rho_1, \quad p=p_0+p_1, \quad s=s_0+s_1, \quad {\cal T}={\cal T}_0 + {\cal T}_1 \, .
\ee
A short computation based on (\ref{th_eq}) shows that $s_1$ and ${\cal T}_1$ are determined in terms of $\rho_1$, as
\be
s_1 = \frac{\rho_1}{{\cal T}_0} ~, ~~~~~~~  \frac{{\cal T}_1}{{\cal T}_0} = \frac{1}{3} \frac{\rho_1}{\rho_0} \, ,
\ee
whereas the correction to the pressure is obtained using $p= \rho /2$.

Using the late time expansion of $T_{ab}$ found in section \ref{late_exp_T} together with the boundary metric 
(\ref{2nd_corr}), we solve $T_{ab} u^b = - \rho u_a$ to find,
\be
\rho_1 = - {16a^2 \over 3m \kappa^2} {\rm sin}^2 \theta
{\rm cos}^2 \theta ~ e^{-4t/m}
\label{kalica}
\ee
The components of the normalized time-like unit vector $u_a$ are given to the same order by
\ba
u_t & = & -1 + {8a^2 \over 3m^2 \Lambda} {\rm sin}^2 \theta
{\rm cos}^2 \theta ~ e^{-4t/m} ~, \\
u_{\theta} & = & {4a \over m \Lambda} {\rm sin} \theta {\rm cos} \theta ~ e^{-2t/m} +
{8a^2 \over 39 m \Lambda} {\rm sin} \theta {\rm cos} \theta \left(2 + 42 {\rm sin}^2
\theta - {39 \over m^2 \Lambda} {\rm sin}^2 \theta \right) e^{-4t/m} ~,  \nonumber  \\
u_{\phi} & = & 0  \nonumber
\ea
and it turns out that there is a non-vanishing divergence
\be
\nabla_a u^a = {8a^2 \over 3m^3 \Lambda} \, {\rm sin}^2 \theta \left(9 {\rm cos}^2 \theta -1 \right)
e^{-4t/m} \,.
\ee
Actually, these particular results hold for all value of $r_{\rm h}$, large or small compared to the AdS radius $L$,
but they will be used next only in the limit of large $r_{\rm h}$.

Using the equilibrium temperature of large AdS black holes ${\cal T}_0$, it follows from \eqn{kalica} that
\be
s_1 = {\rho_1 \over {\cal T}_0} = -{16a^2 \over 9 r_{\rm h}^4} 
\left(-{3 \over \Lambda} \right)^2 {\rm sin}^2 \theta {\rm cos}^2 \theta \, e^{-4t/m} \, .
\ee
We observe that $s_1$ is negative and so the entropy $s$ increases and becomes equal to that of the Schwarzschild solution in the limit
$t \to \infty$, showing that there is entropy production in the problem at hand.
We also compute the divergence of the entropy current, keeping leading order terms in the large $r_{\rm h}$ limit, and find
\be
\nabla_a s^a =  {\partial s_1 \over \partial t} + s_0 \nabla_a u^a =
{16a^2 \over 9 r_{\rm h}^7}  \left(-\frac{3}{\Lambda}\right)^3 \sin^4 \theta \, e^{-4 t/m} \, .
\ee
Thus, the entropy current has a non-negative divergence. Finally, one may check, using equations
(\ref{pi_theta})-(\ref{pi_phi}), that the divergence of the entropy current can be cast in the form
\be
\nabla_a s^a = {\eta \over 2 {\cal T}_0} \, \sigma_{ab} \sigma^{ab} \, ,
\ee
as required on general grounds. This last step makes use of the form for $\sigma_{ab}$ for quadrupole radiation to linear
order, which was determined earlier in section 4.3,
\be
\kappa^2 \eta \sigma_{\theta \theta} = -{6a \over m \Lambda} {\rm sin}^2 \theta \, e^{-2t/m} ~, ~~~~~~
\sigma_{\phi \phi} = - {\rm sin}^2 \theta \, \sigma_{\theta \theta}
\label{maroulia5}
\ee
setting $\kappa^2 \eta = 3/2$ (for $l=2$) and taking proper account of the normalization of $\delta T_{ab}^{(1)}$
(factor of $-4a/3$) appearing in the late time expansion \eqn{laertisa}. The expressions for $\sigma_{ab}$ shown in 
\eqn{maroulia5} are valid
for black holes of all sizes, but we are applying them here only to the case of large $r_{\rm h}$, 
$r_h \gg \sqrt{-3/\Lambda}$ letting $m = - \Lambda r_{\rm h}^3 /6$.

We have obtained the entropy current using local thermodynamics. One may wonder whether this current can also
be obtained directly from a bulk computation, as in \cite{Bhattacharyya:2008xc}. In a time dependent context
one expects the entropy current to be associated with a future apparent horizon \cite{Chesler:2008hg}. Future
apparent horizons lie behind the event horizons and in our case a bulk surface whose area would give the corrected
entropy would also have to lie behind the horizon at some $r<r_h$. However, as we discussed earlier, the
Robinson-Trautman solutions do not extend smoothly across $r=r_h$ and, thus, one does not expect to be able to
geometrize this entropy current.

\section{Conclusions and discussion}
\setcounter{equation}{0}

We discussed various aspects of the $AdS_4$ Robinson-Trautman space-times. This particular class of
solutions may be thought to describe the gravitational field outside a compact source that radiates away its asymmetry
and settles to a spherically symmetric configuration, which is the $AdS_4$ Schwarzschild solution.
The corresponding asymptotically flat solution is well-studied and its global properties are well-known.
In this paper we extended the analysis to the case of negative cosmological constant and also discussed
holographic aspects of the solutions.

There is a number of similarities and differences between Robinson-Trautman metrics with zero and negative
cosmological constant $\Lambda$. In either case there is a naked singularity in the past, as well as a past apparent horizon, which
can be regarded as a surface surrounding the compact source. The evolution is governed by the Calabi flow on $S^2$,
which is a geometric evolution equation that describes certain volume preserving deformations of $S^2$, and it is independent
of $\Lambda$. At late times, the metric on $S^2$ approaches the constant curvature metric and the four-dimensional
Robinson-Trautman solution becomes the Schwarzschild solution. More precise, as the retarded time $u$ tends to infinity,
the future horizon ${\cal H}^+$ of Schwarzschild space-time is reached. The deviation from Schwarzschild metric
at linear order is described by algebraically special modes. Remarkably, these modes are supersymmetric zero energy states
of the effective quantum mechanics problem for the linear perturbations of the Schwarzschild solution.

The algebraically special modes describe out-going gravitational radiation and they vanish at ${\cal H}^+$.
One may ask whether these modes, as well as the Robinson-Trautman solution, admit a Kruskal extension across the horizon.
The result depends on the cosmological constant. When $\Lambda = 0$, all individual modes have a smooth extension,
but this is severely affected by non-linear effects. When $\Lambda < 0$, the low lying multi-pole modes of large
$AdS_4$ Schwarzschild black holes do not appear to have a Kruskal extension at all. The curvature invariants are smooth at
${\cal H}^+$, but it is expected that at least one of the component of  (covariant derivatives of) the curvature tensor blows
up at ${\cal H}^+$ in all smooth coordinate systems that extend through ${\cal H}^+$.
This indicates that there is a null singularity at the future horizon.

We also considered the Bondi mass, which is a characteristic physical quantity of radiative solutions that decreases monotonically
at late times. We proved that it satisfies
a Penrose inequality and formulated and provided evidence for a version of the hoop conjecture for $\Lambda <0$.
While the Bondi mass is a natural concept in asymptotically flat gravity, measuring the mass at null infinity,
its meaning is less clear with AdS asymptotics, since there is no null infinity in that case. In this paper, we assigned a Bondi mass
to the $AdS_4$ Robinson-Trautman space-times in exact analogy with the $\Lambda=0$ case. It would be interesting to understand
whether the analogue of Bondi mass can be defined in general asymptotically locally AdS space-times from first principles.

The space-time has a holographic interpretation when the cosmological constant is negative. In that context,
the relaxation to $AdS_4$ Schwarzschild describes the approach to equilibrium in the dual QFT. One of the basic observables
is the holographic energy-momentum tensor $T_{ab}$. It encodes the ADM conserved charges from the bulk perspective
and it computes the expectation value of the energy and momentum in the system from the dual QFT perspective.
Near thermal equilibrium, $T_{ab}$ takes the form of dissipative hydrodynamics. It turns out that the effective shear viscosity
is not universal and for low multi-poles the KSS bound is violated. The modes, however, are out-going rather than in-coming at the horizon,
they do not satisfy Dirichlet boundary conditions at the conformal boundary and they do not have a Kruskal extension
across the future horizon. In that case, dissipation is due to the coupling to external sources. Thus, the physics of the system
is very different from those studied in the literature (it is well known that the KSS bound can be violated by perturbations
that become singular at the horizon, but the nature of the singularity is different here).

It would be interesting to understand better the gauge/gravity duality of these solutions, focusing on the meaning of the
unusual boundary condition imposed by the algebraically special modes. It would also be interesting to understand the
meaning of the Bondi mass from the perspective of the dual QFT as well as the boundary manifestation of the Penrose inequality and
the hoop conjecture. Finally, it would be interesting to investigate whether one can produce a smooth metric by joining the
Robinson-Trautman solution to matter-filled interior. This could lead to the exciting possibility of describing holographically
compact sources. We hope to return to these problems elsewhere.

\vskip0.5cm

{\bf Note Added:} While this paper was in final states we received \cite{reall} which contains related material.
In particular, the authors analyze the time-reversed Robinson-Trautman solution and present a thorough analysis of hydrodynamics. Their results for the energy-momentum tensor of Robinson-Trautman and the boundary Cotton tensor (after taken into account the time reversal) are identical to ours.

\vskip0.7cm

\centerline{\bf Acknowledgements}
\noindent
This research of I.B. is partially supported and implemented under the
ARISTEIA action of the operational programme for education and long life learning
and is co-funded by the European Union (European Social Fund) and
National Resources of Greece. K.S. acknowledges support from a grant of the John Templeton
Foundation. The opinions expressed in this publication are those of the authors and do not
necessarily reflect the views of the John Templeton Foundation. Preliminary versions of this
work were presented in a number of workshops, including the ``New frontiers in
dynamical gravity" held 24-28 of March 2014 at Cambridge, UK.

\newpage

\appendix
\section{Derivation of Robinson-Trautman equation}
\setcounter{equation}{0}

The Robinson-Trautman metric (1.1) with front factor $F$ given by (1.2) involves
a parameter $m$, which, in general, can be taken to depend upon $u$. Then,
the Robinson-Trautman equation assumes the following general form, irrespective of the
cosmological constant $\Lambda$, which can be positive, negative or zero,
\be
3m(u) \partial_u \Phi + \Delta \Delta \Phi + 2 \partial_u m(u) = 0 ~.
\label{rt}
\ee

The derivation is done by successive integration of Einstein equations $R_{\mu \nu}
= \Lambda g_{\mu \nu}$ after computing the components of the Ricci curvature tensor,
\ba
R_{uz} & = & {1 \over 2} \partial_z \left(\partial_r F - \partial_u
\Phi \right) , \\
R_{u\bar{z}} & = & {1 \over 2} \partial_{\bar{z}} \left(\partial_r F
- \partial_u \Phi \right) , \\
R_{z \bar{z}} & = & 2re^{\Phi} (\partial_u \Phi) - \partial_z
\partial_{\bar{z}} \Phi - e^{\Phi} \partial_r(rF) ~, \\
R_{ur} & = & {1 \over 2} \partial_r^2 F + {1 \over r} \left(\partial_r F
- \partial_u \Phi \right) , \\
R_{uu} & = & - \partial_u^2 \Phi - {1 \over 2} (\partial_u \Phi)^2 +
{1 \over 2} F \left(\partial_r^2 F + {2 \over r} \partial_r F \right) \nonumber\\
& & -{1 \over 2} (\partial_r F) (\partial_u \Phi) + {1 \over r}
\partial_u F + {1 \over r^2} \Delta F ~,
\ea
whereas all other components vanish identically.

First, we integrate the $(uz)$ and $(u \bar{z})$ components
of Einstein equations and obtain that $\partial_r F - \partial_u \Phi$
is independent of $z$ and $\bar{z}$. Thus, we have
\be
\partial_r F - \partial_u \Phi = \lambda (r, u) ~,
\ee
where $\lambda (r, u)$ will be determined shortly.
Next, we integrate the $(z \bar{z})$ component of Einstein
equations and use the previous result to determine $F$ as follows,
\be
F = r(\partial_u \Phi) - \Delta \Phi - r \lambda (r, u) - \Lambda r^2 ~.
\ee
Taking the above into account, the $(ur)$ component of Einstein
equations yields a differential equation for $\lambda (r, u)$,
\be
\partial_r^2 \left(r \lambda (r, u) \right) = {2 \over r} \lambda (r, u) ~,
\ee
whose general solution is
\be
\lambda (r, u) = \alpha (u) r + {\beta (u) \over r^2} ~.
\ee
Finally, using the $(uu)$ component of Einstein equations, we
arrive after some manipulation to the following equation,
\be
\Big[\partial_u \alpha (u) + (\partial_u \Phi) \left(\Lambda +
{3 \over 2} \alpha (u) \right) \Big] r +
\Big[\partial_u \beta (u) + \Delta \Delta \Phi + {3 \over 2}
\beta (u) (\partial_u \Phi) \Big] {1 \over r^2} = 0 ~,
\ee
which is valid for all $r$. Thus, the coefficients of the $r$ and $1/r^2$
should vanish separately and one obtains, respectively, the system of equations
\ba
& & \partial_u \alpha (u) + (\partial_u \Phi) \left(\Lambda +
{3 \over 2} \alpha (u) \right) = 0 ~, \\
& & \partial_u \beta (u) + \Delta \Delta \Phi + {3 \over 2}
\beta (u) (\partial_u \Phi) = 0 ~.
\ea
Since $\Phi$ depends upon $z$ and $\bar{z}$, apart from $u$, the first equation implies
automatically that
\be
\alpha (u) = -{2 \over 3} \Lambda ~,
\ee
whereas the second equation, setting
\be
\beta (u) = 2 m(u) ~,
\ee
implies the Robinson-Trautman equation \eqn{rt} with $u$-dependent mass term $m$, in
general.

Note, however, that the Robinson-Trautman line element, as we have it so far, is form-invariant
under the following class of coordinate transformations,
\be
u^{\prime} = \rho (u) ~, ~~~~ r^{\prime} = {r \over \partial_u \rho (u)} ~,
~~~~ z^{\prime} = f(z) ~, ~~~~ \bar{z}^{\prime} = \bar{f}(\bar{z})
\ee
provided that
\be
e^{\Phi^{\prime} (z^{\prime} , \bar{z}^{\prime} ; u^{\prime})} =
{(\partial_u \rho(u))^2 \over \mid \partial f \mid^2}
e^{\Phi (z, \bar{z}; u)}
\ee
and
\be
m^{\prime} (u^{\prime}) = {m(u) \over (\partial_u \rho (u))^3} ~.
\ee
Thus, by appropriate choice of $\rho (u)$ one can fix $m(u)$ to a constant
and transform the RT equation to the familiar form. This
analysis also shows how $\Phi (z, \bar{z}; u)$ should transform under changes of $z$ and
$\bar{z}$ coordinates.

\section{Large and small $AdS_4$ black holes}
\setcounter{equation}{0}

Einstein equations in four space-time dimensions with cosmological constant
$\Lambda$ admit the Schwarzschild metric as spherically symmetric static solution,
\be
ds^2 = -f(r) dt^2 + {dr^2 \over f(r)} + r^2 \left(d\theta^2 +
{\rm sin}^2 \theta d\phi^2 \right) ,
\ee
with profile function
\be
f(r) = 1 - {2m \over r} - {\Lambda \over 3} r^2
\ee
that gives the appropriate asymptotic behavior, as $r \rightarrow \infty$, for all values of $\Lambda$.

Here, we mainly consider $AdS_4$ Schwarzschild metric, having $\Lambda <0$.
The Schwarzschild radius of $AdS_4$ black holes is provided by the real root of $f(r) = 0$ occurring at
\be
r_{\rm h} = {1 \over \sqrt{-\Lambda}} ~
\big[\left(\sqrt{1 - 9m^2 \Lambda}
+ 3m \sqrt{-\Lambda} ~ \right)^{1/3} -
\left(\sqrt{1 - 9m^2 \Lambda}
- 3m \sqrt{-\Lambda} ~ \right)^{1/3} \big] ~.
\ee
Thus, the black hole radius takes values $0 < r_{\rm h} < 2m$ depending
on the size of $\Lambda$. When $\Lambda$ approaches zero, $r_{\rm h}$ tends
to $2m$, whereas for $\Lambda << 0$, $r_{\rm h}$ comes close to 0.

It is also useful to introduce the {\em tortoise} coordinate $r_{\star}$, which
is defined by
\be
dr_{\star} = {dr \over f(r)} ~.
\ee
When $\Lambda = 0$, $r_{\star}$ ranges from $-\infty$ to $+\infty$
as $r$ ranges from the black hole horizon located at $r=r_{\rm h}$ to
infinity. But when $\Lambda < 0$, which is the case of interest here, $r_{\star}$
ranges from $-\infty$ up to some finite value that can be set equal to
zero by appropriate choice of integration constant. For $AdS_4$ black
holes, in particular, we have explicitly
\be
r_{\star}  =  {r_{\rm h} \over 4 (r_{\rm h} - 3m)}
\left(r_{\rm h} ~ {\rm log} {(2r + r_{\rm h})^2 + a^2  \over
4 (r-r_{\rm h})^2} +
2 a ~ {r_{\rm h} - 6m \over
r_{\rm h} + 6m} ~ \big[{\rm arctan}
{2r + r_{\rm h} \over a} -{\pi \over 2} \big]\right) ,
\ee
setting for notational convenience
\be
a = \sqrt{-{3 \over \Lambda} \left(1 + {6m \over r_{\rm h}} \right)} ~ .
\ee

AdS black holes come in different sizes and their thermodynamic properties
depend crucially on the magnitude of $r_{\rm h}$ relative to the AdS radius
$L = \sqrt{-3 / \Lambda}$, \cite{don}.
Large black holes have $r_{\rm h} > L$ and become the dominant configurations
at high temperatures, whereas small black holes have $r_{\rm h} < L$ and they
are always unstable decaying either to pure thermal radiation or to black
holes with larger mass. In general, we have the following relation among the
parameters of the $AdS_4$ Schwarzschild background
\be
m - r_{\rm h} = {1 \over 2L^2} r_{\rm h} \left(r_{\rm h}^2 - L^2 \right) .
\ee
Thus, large black holes have $r_{\rm h} < m$, whereas small black holes have
$r_{\rm h} > m$.

Finally, we recall that very large black holes are naturally
associated to the limit $r_{\rm h} \rightarrow \infty$, in which case
$f(r)$ is simply replaced by
\be
f(r) = -{2m \over r} - {\Lambda \over 3} r^2 ~,
\ee
dropping the constant term. Then, the black holes become essentially flat
and their horizon is related to the other parameters by the simple
expression
\be
r_{\rm h}^3 = - {6m \over \Lambda} ~.
\ee

\section{Polar perturbations of spherical black holes}
\setcounter{equation}{0}

This is a particular class of metric perturbations of Schwarzschild solution (in the presence of
cosmological constant), which have the parity of Legendre polynomial $P_l({\rm cos} \theta)$, i.e.,
$(-1)^l$. For simplicity, we only consider axisymmetric perturbations, which are parametrized by four
arbitrary radial functions and they assume the general form
\be
\delta g_{\mu \nu} = \left(\begin{array}{cccc}
f(r)H_0(r) & H_1(r) & 0 & 0 \\
  &   &   &   \\
H_1(r) & H_2(r)/f(r) & 0 & 0 \\
  &   &   &   \\
0 & 0 & r^2K(r) & 0 \\
  &   &   &   \\
0 & 0 & 0 & r^2K(r) {\rm sin}^2 \theta
\end{array} \right)
e^{-i\omega t}
P_l ({\rm cos} \theta) ~.
\ee
They correspond to the so called
sound modes in the dictionary of AdS/CFT correspondence. As such, they should be
contrasted to the axial perturbations of spherical black holes, also called shear modes in the
context of AdS/CFT, which form a complementary set of perturbations with opposite parity, i.e.,
$(-1)^{l+1}$, but their explicit form is not needed for the purposes of the present work. We refer
the reader to the original papers \cite{wheeler, zerilli}, the textbook \cite{chandra2}, the
review article \cite{kokko} as well as the more recent papers \cite{lemos, moss, Bakas:2008gz} for
generalizations in the presence of cosmological constant.

We require $l \geq 2$ to describe physical modes of gravitational radiation in the
linearized approximation (recall that there is no dipole radiation in general relativity).
The study of polar perturbations $\delta R_{\mu \nu} = \Lambda \delta g_{\mu \nu}$ about the
Schwarzschild background yields
\be
H_0(r) = H_2(r) ~.
\ee
Tedious computation also shows that the $(tr)$ $(r\theta)$ and
$(t\theta)$ components of the perturbation yield the following
equations, respectively,
\ba
& & r K^{\prime}(r) + \left(1 - {rf^{\prime}(r) \over 2f(r)}
\right) K(r) - H_0(r) - i {l(l+1) \over 2 \omega r} H_1(r) = 0 ~, \\
& & \left(f(r) H_0(r)\right)^{\prime} - f(r) K^{\prime}(r)
+ i\omega H_1(r) = 0 ~, \\
& & \left(f(r) H_1(r)\right)^{\prime} + i\omega \left(H_0(r)
+ K(r) \right) = 0 ~.
\ea
Together they form a coupled system of first order differential
equations for the three unknown functions $H_0(r)$, $H_1(r)$ and $K(r)$.
The other components of the perturbation either yield second order equations or
else $\delta R_{\mu \nu}$ vanishes identically.
Note, however, that there is an additional algebraic condition among the
three radial functions
\ba
& & \left(2f(r) - rf^{\prime}(r) - l(l+1) \right) H_0(r) +
{i \over 2\omega} \left(4\omega^2 r - l(l+1) f^{\prime}(r) \right)
H_1(r) = \nonumber \\
& & \left(2f(r) + rf^{\prime}(r) - l(l+1) + 2\Lambda r^2 +
{r^2 \over 2f(r)} \left(4 \omega^2 + {f^{\prime}}^2(r) \right)
\right) K(r) ~,
\ea
which follows from consistency of the various second order equations with
the first order system above; it can also be viewed as integral of the
first order system above.

Following Zerilli, \cite{zerilli}, we define the following function,
\be
\Psi_{+} (r) = {r^2 \over (l-1)(l+2)r + 6m} \left(K(r)
- i{f(r) \over \omega r} H_1 (r) \right) ,
\ee
which turns out to satisfy an effective Schr\"odinger equation with respect to the tortoise
coordinate $r_{\star}$,
\be
\left(-{d^2 \over dr_{\star}^2} + V_{+}(r) \right) \Psi_{+}(r)
= \omega^2 \Psi_{+}(r) ~,
\ee
where the potential is only determined implicitly in terms of $r_{\star}$ via
\ba
V_{+}(r) & = & {f(r) \over [(l-1)(l+2) r + 6m]^2}
\left(l(l+1)(l-1)^2 (l+2)^2 - 24m^2 \Lambda \right. \nonumber \\
& & \left. + {6m \over r} (l-1)^2 (l+2)^2
+ {36m^2 \over r^2}(l-1)(l+2)
+ {72 m^3 \over r^3} \right) .
\ea
The solutions determine the quasi-normal mode
spectrum $\omega$ of the polar perturbations under appropriate boundary conditions and
lead to expressions for the three unknown radial functions described above.

It turns out that the Zerilli potential admits the following representation
\be
V_{+}(r) = W^2(r) + {d W(r) \over dr_{\star}} + \omega_{\rm s}^2
\ee
in terms of the following (real) function
\be
W(r) = {6m f(r) \over r[(l-1)(l+2)r + 6m]} + i \omega_{\rm s} ~,
\ee
where
\be
\omega_{\rm s} = -{i \over 12m} (l-1)l(l+1)(l+2) ~.
\ee

There is parallel story to be told about the parity odd (axial) perturbations, but we skip details that
are not relevant to the present work. In that case, the important thing is that there is an effective
Schr\"{o}dindger problem for the perturbations with corresponding potential,
called Regge-Wheeler potential, \cite{wheeler}, which is also written in terms of $W$ as
\be
V_{-}(r) = W^2(r) - {d W(r) \over dr_{\star}} + \omega_{\rm s}^2 ~.
\ee
Thus, black hole perturbations in four space-time dimensions are neatly described by two complementary
Schr\"odinger problems, for all $\Lambda$,
\be \label{twoschrod}
H_{\pm} \Psi_{\pm} = E \Psi_{\pm} \, ,
\ee
where the Hamiltonian are given by
\be
H_{\pm}=-{d^2 \over dr_{\star}^2} + W^2 \pm {dW \over dr_{\star}} ~.
\ee
The energy $E = \omega^2 - \omega_{\rm s}^2$ is not bounded from below by zero (in fact, it is not even real,
in general) due to the particular boundary conditions imposed on the wave functions at the black hole horizon.

Remarkably, the system (\ref{twoschrod}) has the structure of supersymmetric quantum mechanics (see, for instance, \cite{susy}).
Introducing the first order operators (analogue of annihilation and creation operators of a
harmonic oscillator)
 \be \label{qdef}
Q = -{d \over d r_{\star}} + W(r_{\star}) ~, ~~~~~~
Q^{\dagger} = {d \over d r_{\star}} + W(r_{\star}) ~,
\ee
we have a pair of Hamiltonians
\be
H_{+} = Q^{\dagger} Q = -{d^2 \over dr_{\star}^2} + W^2 + {dW \over dr_{\star}} ~, ~~~~~~
H_{-} = Q Q^{\dagger} = -{d^2 \over dr_{\star}^2} + W^2 - {dW \over dr_{\star}}
\ee
corresponding to the two Schr\"odinger problems \eqn{twoschrod}. They give rise to a supersymmetry
algebra,
\be
\{{\cal Q} , ~ {\cal Q}^{\dagger} \} = {\cal H} \, , \qquad \{{\cal Q} , ~ {\cal Q} \}= 0 =
\{{\cal Q}^{\dagger} , ~ {\cal Q}^{\dagger} \} \, ,
\ee
in terms of supercharges
\be \label{Qdef}
{\cal Q} = \left(\begin{array}{ccc}
0 &  & 0 \\
  &   & \\
Q &  & 0
\end{array} \right) , ~~~~~~
{\cal Q}^{\dagger} = \left(\begin{array}{ccc}
0 &  & Q^{\dagger} \\
  &   & \\
0 &  & 0
\end{array} \right)
\ee
and
\be
{\cal H} = \left(\begin{array}{ccc}
H_{+} &  & 0 \\
  &   & \\
0 &  & H_{-}
\end{array} \right) .
\ee
Defining
\be
\Psi = \left(\begin{array}{c}
\Psi_+ \\
\Psi_- \\
\end{array} \right) ,
\ee
the two Schr\"odinger problems \eqn{twoschrod} take the form
\be
{\cal H} \Psi = E \Psi ~.
\ee
The Hamiltonian is only formally Hermitian because of the physical boundary conditions that give rise
to the quasi-normal mode spectrum of the black hole.

The algebraically special modes, which are discussed in the text, are special case
of polar perturbations, satisfying the first order equation
\be
Q \Psi (r_{\star}) = \left(-{d \over d r_{\star}} + W(r_{\star}) \right) \Psi (r_{\star}) = 0 ~.
\ee
As such, they are supersymmetric zero energy states of $H_+$ having no partner in the
complementary set of axial perturbations.

It is a mystery up to this day why the two complementary sets of perturbations of four-dimensional black holes
form supersymmetric partner potentials although the space-time itself has no supersymmetry. This
observation, which was first made by Chandrasekhar for $\Lambda = 0$ \cite{chandra2} (but see also earlier references
therein) and subsequently generalized to $\Lambda \neq 0$ (see, for instance, \cite{Bakas:2008gz}).

\end{document}